\documentclass[aps,prb,twocolumn,groupedaddress,showpacs,floatfix,superscriptaddress,longbibliography]{revtex4-2}
\usepackage[plainpages=false,pdfpagelabels,colorlinks=true,linkcolor=red,urlcolor=blue,citecolor=blue,pdftitle={Title},pdfauthor={Rubem Mondaini},pdfdisplaydoctitle=true,pdfduplex=DuplexFlipLongEdge]{hyperref}
\usepackage{epsfig}
\usepackage{amsmath,amssymb}
\usepackage{physics}
\usepackage{graphicx}
\usepackage[dvipsnames,usenames]{color}
\usepackage[normalem]{ulem}
\usepackage{soul} 
\usepackage{float}
\usepackage{diagbox}
\usepackage{orcidlink}

\newcommand{\tcr}{\textcolor{red}}

\DeclareRobustCommand{\tcr}[1]{\textcolor{black}{#1}}

\begin{document}
\title{Photodynamic melting of phase-reversed charge stripes and enhanced condensation}

\author{Jianhao Sun}
\affiliation{Beijing Computational Science Research Center, Beijing 100193, China}

\author{Richard T.~Scalettar\,\orcidlink{0000-0002-0521-3692}}
\email{scalettar@physics.ucdavis.edu}
\affiliation{Department of Physics and Astronomy, University of California, Davis, CA 95616, USA}

\author{Rubem Mondaini\,\orcidlink{0000-0001-8005-2297}}
\email{rmondaini@uh.edu}
\affiliation{Department of Physics, University of Houston, Houston, Texas 77004, USA}
\affiliation{Texas Center for Superconductivity, University of Houston, Houston, Texas 77004, USA}

\begin{abstract}
The interplay between charge stripes and pairing has long been a subject of scrutiny in a broad class of unconventional superconductors, as in some cases it is unclear whether this interplay benefits the ensuing superfluidity. Experiments that explore the out-of-equilibrium dynamics of these systems aim to tip the balance toward one phase or the other by selectively coupling to relevant modes. Leveraging the fact that competition between stripes and pairing is not exclusive to fermionic systems, we explore the photoirradiation dynamics of interacting hardcore bosons, in which density-wave phase-reversal melting leads to enhanced \tcr{phase-coherent transport response}, as quantified by the dynamic amplification of \tcr{both the} zero-momentum occupancy \tcr{and the condensate fraction}, \tcr{as well as finite out-of-equilibrium} charge stiffness \tcr{and superfluid weight, for a given system size}. Our results, obtained using unbiased methods for an interacting system on a ladder geometry, demonstrate how one can engineer time-dependent perturbations to release suppressed orders, potentially providing insight into the underlying mechanism in related experiments.
\end{abstract}

\maketitle

\section{Introduction} Time-dependent experiments provide another angle in the longstanding quest to understand the mechanism binding otherwise repulsive fermions, supporting the emergence of unconventional superconductivity. In particular, rapid energy insertion in the form of an ultrafast laser pump has arisen as a capable technique~\cite{Zhang2014, Cavalleri2018} that coherently drives the dynamics of the electrons in ways that can control magnetic~\cite{Kirilyuk2010, Siegrist2019}, charge~\cite{Kogar2020, Wandel2022, Tuniz2023, Duncan2025}, and superconducting orders~\cite{Fausti2011, Hu2014, Nicoletti2014, Kaiser2014, Liu2024, Fava2024, Nishida2024}. Owing to the entwining of the different orders that permeate the doped cuprate physics~\cite{Fradkin2015, Fradkin2025}, such an approach is one in which coupling to certain degrees of freedom can suppress specific ordering while releasing others~\cite{Cavalleri2018, Wang2018}.

For instance, in equilibrium, the presence of charge stripes, quasi-one-dimensional regions of doped charges (either holes or electrons), is ubiquitous in a large class of cuprate superconductors, often accompanied by phase reversal of local order, such as antiferromagnetic spin alignment~\cite{Keimer2015, Devereaux2025}. Evidence abounds that when such density modulation is static, it directly hurts superconductivity, as in the case of La$_{2-x}$Ba$_{x}$CuO$_4$ cuprates~\cite{Tranquada1995}, in particular at  $x=1/8$-doping~\cite{Fujita2002, Abbamonte2005, Tranquada2008, Tranquada2020}. Nonetheless, for many other compounds, including bismuth-~\cite{Parker2010, Neto2014, Chaix2017} and yttrium-based cuprates~\cite{Wu2011, Ghiringhelli2012, Chang2012, Blanco-Canosa2013}, fluctuating stripes associated with incommensurate ordering wave-vectors have been observed, and these often can coexist with coherent pairing.

Although the physics of stripes in equilibrium, including fluctuating/incommensurate ones, is now well studied within a broad class of models and numerical methods~\cite{Mondaini2012, Zheng2017, Huang2017, Huang2018, Jiang2019, Wietek2021, Xiao2023, Xu2024, Chen2024, Liu2025}, direct indication of the \textit{dynamical} interplay between charge ordering and pairing is nonexistent in unbiased numerics of many-body models even though this is the precise mechanism explored in experiments~\cite{Fausti2011, Forst2014, Nicoletti2014, Cremin2019} -- we aim to fill this void~\footnote{We note microscopic models for cuprates have been studied under photoirradiation~\cite{Wang2020, Tang2021, Wang2021, Wang2021b}, but the interplay with stripe physics is currently missing.}. Our key observation is that the stripe physics and antiphase of local order are not intrinsic to repulsive interactions~\cite{Ying2022}, nor to fermionic statistics~\cite{Sun2024}, including the fluctuating-static dichotomy~\cite{Sun2024}. This fundamental, and highly non-trivial, analogy allows one to unbiasedly solve the dynamics of Hamiltonians that contain specific ingredients of actual cuprate physics --- namely, strong interactions, superfluidity, and antiphase correlations across stripes --- even if the pairing is isotropic, unlike what is observed in a large class of unconventional superconductors, and even though our model does not contain the spin degrees of freedom which are relevant for cuprate materials.

Despite these differences,
our main result is that by adjusting the characteristics of the time-dependent perturbation in the form of a pump in a strongly correlated system, one can show that an effective nonlinear optical coupling is responsible for exciting the system from the ground state in a way that a static stripe with antiphase ordering melts \tcr{while coherence and dissipationless-transport diagnostics are enhanced dynamically}, a feature in line with experimental exploration of actual cuprates under photoirradiation.

\section{Model} 
\begin{figure*}[!t]
    \centering
    \includegraphics[width=1\linewidth]{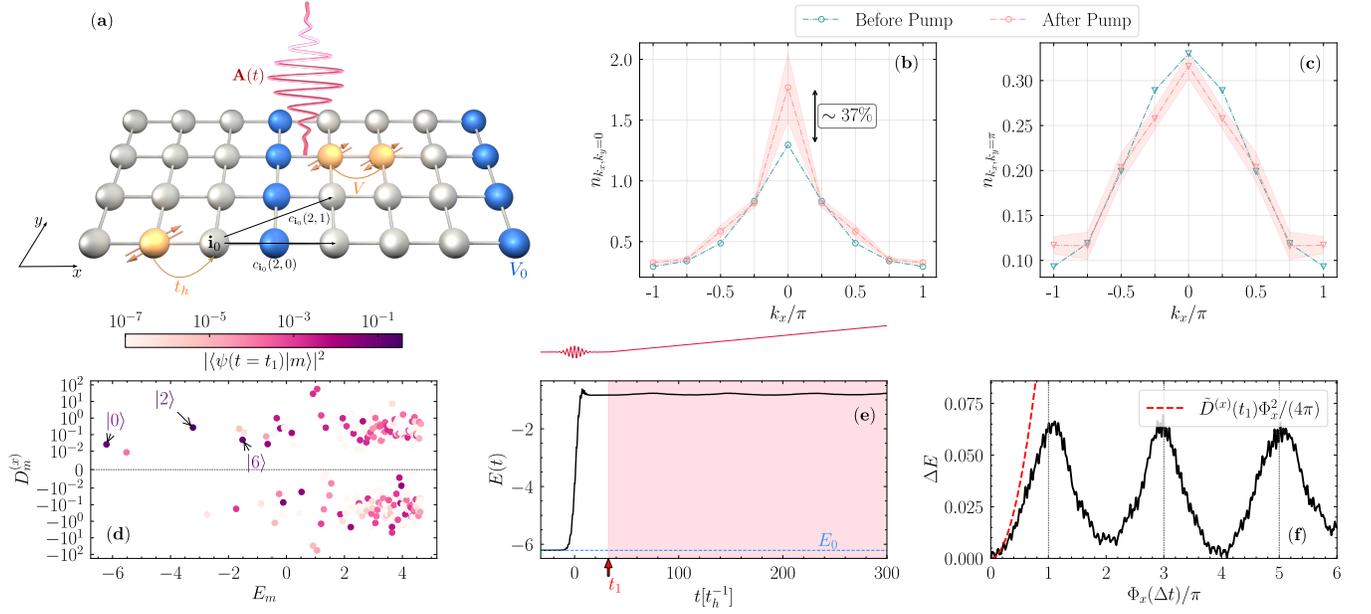}
    \caption{(a) Schematic illustration of Eq.~\eqref{fig:fig1} with relevant terms annotated, subjected to $x$-polarized photoirradiation. Here, the hardcore bosons are represented by a composite local fermion. Momentum distribution profiles $n_{\bf k}$ focusing on $k_y = 0$ (b) and $k_y = \pi$ (c), comparing the equilibrium (before pump) with the out-of-equilibrium (after pump, averaged results from $t \in [11t_d, 26t_d] = [72, 169]t_h^{-1}$; shading gives the standard deviation) response to the external field. We observe an average relative enhancement of about 37\% for the zero-momentum condensation under the photoirradiation. (d) The $x$-direction charge stiffness resolved for the first 200 eigenstates $|m\rangle$ of $\hat {\cal H}$; marker colors are mapped by the overlap of the time evolved state at $t_1=5t_d$ and the corresponding $|m\rangle$. (e) The instantaneous energy $E(t) = \langle \psi(t)|\hat {\cal H}(t)|\psi(t)\rangle$ over the dynamics; the shaded region marks the probing for the dynamical charge stiffness starting at $t_1$, with a linear-in-time vector potential with slope $\delta A_x$ --- top panel depicts the total ${\bf A}(t)$. Inset shows the energy variation in the probe regime as a function of the effective flux $\Phi_x(t)=\delta A_x L_x(t-t_1)$ for $t>t_1$;\ $\delta A_x=0.01$. Dashed line displays an approximant at small fluxes that quantifies the dynamical charge stiffness at $t =t_1$, confirming a robust charge transport induced by the pump (see text). (f) The dynamics of the condensate fraction, highlighting a $\sim 34\%$ increase when considering the running average (darker curve). Pulse parameters are $\Omega = 1.713\ t_h$, $A_0 = 0.62$ and $t_d = 6.5 t_h^{-1}$.}
    \label{fig:fig1}
\end{figure*}

Our starting point is the Hubbard model for repulsive hardcore bosons, 
\begin{align}
    \hat {\cal H} =-t_h\sum_{\langle {\bf i,j} \rangle }\hat b_{\bf i}^\dagger \hat b_{\bf j}^{\phantom{\dagger}}
     + V \sum_{\langle {\bf i,j} \rangle } \hat n_{\bf i} \hat n_{\bf j}
    + V_0 \sum_{{\bf i}\in \mathcal P} \hat n_{\bf i} \ ,
    \label{eq:model}
\end{align}
where $\hat b_{\bf i}^{\phantom{\dagger}}$ ($\hat b_{\bf i}^\dagger$) annihilates (creates) a hardcore boson at site ${\bf i}$,  satisfying the statistics $[\hat b_{\bf i}^{\phantom{\dagger}}, \hat b_{\bf j}^\dagger]=0$ if ${\bf i}\neq {\bf j}$ and $\{\hat b_{\bf i}^{\phantom{\dagger}},\hat b_{\bf i}^\dagger\}=1$\footnote{The hardcore constraint is further imposed by $(\hat b_{\bf i}^\dagger)^2=0$ and $(\hat b_{\bf i}^{\phantom{\dagger}})^2=0$}, of an $L_x\times L_y$ ladder with periodic boundary conditions -- we focus on the case $L_x = 2L_y=8$ [Fig.~\ref{fig:fig1}(a)]; $\hat n_{\bf i} = \hat b_{\bf i}^\dagger \hat b_{\bf i}^{\phantom{\dagger}}$ is the associated number operator. The nearest-neighbor, $\langle {\bf i},{\bf j}\rangle$, hopping energy scale is given by $t_h$, and a contact interaction proportional to $V$ establishes the repulsion among the particles. These are supplemented by a modulated chemical potential of amplitude $V_0$, constrained to obey $\mod{(i_x, \mathcal P)} \equiv 0$, where $\mathcal{P}=4$, as a way to induce static stripe formation explicitly with periodicity ${\cal P}$ -- see Fig.~\ref{fig:fig1}(a). While unnecessary in larger ladders~\cite{Sun2024}, it aids in triggering a density modulation characteristic of stripe physics in a relatively small system size.

The dynamics under photoirradiation are emulated by including a time-dependent vector potential, ${\bf A}(t)$. As with fermions, hardcore bosons couple to ${\bf A}(t)$, provided they have an associated charge, through the Peierls' substitution, $\hat b^{\dagger}_{{\bf i}} \hat b^{\phantom{\dagger}}_{{\bf j}} \rightarrow e^{i\mathbf{A}(t) \cdot (\mathbf{i}-\mathbf{j})}\, \hat b^{\dagger}_{{\bf i}} \hat b^{\phantom{\dagger}}_{{\bf j}}$. Focusing on ultrafast pumps, the light pulse is modeled as  ${\bf A}(t)=A_0 e^{-t^2/2t_d^2} \cos( \Omega t +\phi_t ) {\bf e}_{\rm pol}$; assigning a temporal width $t_d$, amplitude $A_0$, frequency $\Omega$, and $\phi_t$ is a temporal phase shift. The polarization direction is chosen as ${\bf e}_{\mathrm{pol}} = \hat {\bf x}$, i.e., perpendicular to the stripes --- see Fig.~\ref{fig:fig1}(a). The unitary dynamics, $|\psi(t+dt)\rangle = e^{-{i}\hat{\mathcal{H}}(t)dt}|\psi(t)\rangle$ [$\hbar \equiv 1$], is computed via Krylov subspace methods~\cite{petsc, slepc}, by taking $|\psi(t\to-\infty)\rangle = |0\rangle$, the equilibrium ground-state of \eqref{eq:model} in the zero quasi-momentum sector where it resides for our set of parameters~\footnote{Since the pump does not break translation invariance, this is the sector the dynamics explore.}.

\section{Enhanced condensation and dissipationless transport} 
\begin{figure}[t]
    \centering
    \includegraphics[width=1\linewidth]{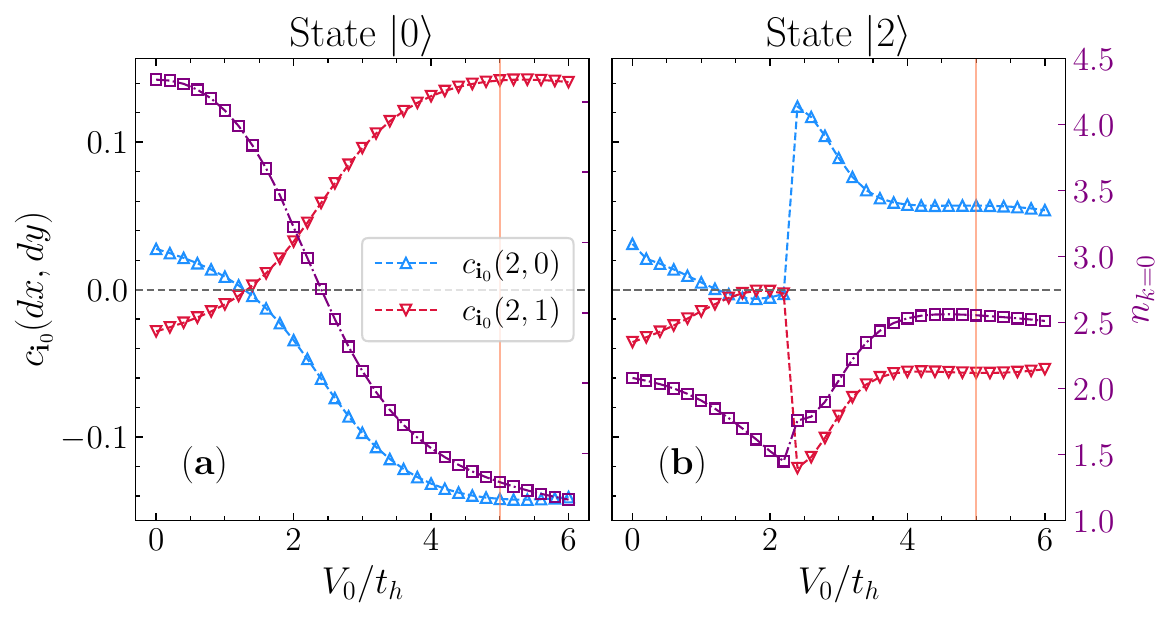}
    \caption{Dependence of the density-density correlations across a stripe for the ground state (a) and the second excited state (b) on the stripe potential. The solid vertical line indicates our selected parameter set for the dynamics, $V_0/t_h=5$, featuring a robust charge $\pi$-phase shift for the ground state, which is absent in the excited state. The right vertical axes display the zero-momentum occupancies.} 
    \label{fig:fig2}
\end{figure}

Interacting hardcore bosons at $T=0$ give rise to either charge ordered phases or a superfluid phase~\cite{Scalettar1995, Hebert2001, Schmid2002, Sengupta2005, Wessel2005, Zhu2020}, depending on the interaction-hopping ratio $V/t_h$ --- in what follows we set $V/t_h=4$. Tendency towards phase separation and stripe formation in the vicinity of half-filling, $\langle \hat n\rangle=1/2$, has been reported~\cite{Sengupta2005, Zhu2020}, and stabilization of such a regime can be made explicit via an external periodic potential~\cite{Sun2024} in the doped regime~\footnote{Including an extra potential has been used in other contexts to stabilize magnetic ordering for example~\cite{Assaad2013}}. Notably, in this scenario, the propensity for spatial organization of holes is accompanied by a reversal in the phase of the checkerboard density modulation across a stripe boundary provided the corresponding chemical potential $V_0$ is sufficiently large -- this is often dubbed as a $\pi$-phase shift~\cite{Mondaini2012, Ying2022, Sun2024}. Focusing on the hole density $\delta \equiv 1/2 - \langle \hat n\rangle=1/8$, Fig.~\ref{fig:fig2}(a) displays the connected density-density correlations across a stripe, $c_{{\bf i}_0}(dx,dy) = \langle \hat{n}_{{\bf i}_0} \hat{n}_{{\bf i}+d{\bf r}}\rangle - \langle \hat{n}_{{\bf i}_0}\rangle\langle\hat{n}_{{\bf i}+d{\bf r}}\rangle$; ${\bf i}_0$ is a reference site and $d{\bf r}\equiv(dx,dy)$ the spatial displacement. As $V_0/t_h$ increases, the same sublattice correlation $c_{{\bf i}_0}(2,0)$ across a stripe transitions from positive to negative row whereas $c_{{\bf i}_0}(2,1)$ shows the opposite trend, establishing a direct verification of the antiphase charge ordered pattern once hole stripes are sufficiently robust.

At the same time, a measure of (quasi-)condensation is given by the momentum distribution function $n_{{\bf k}} = \frac{1}{L_xL_y}\sum_{{\bf i},{\bf j}} e^{-i{\bf k}\cdot({\bf i}-{\bf j})}\langle \hat{b}^\dagger_{\bf i} \hat{b}_{\bf j}^{\phantom{\dagger}} \rangle$, which, at zero temperature, exhibits dependence that is extensive in the system size for the zero-momentum occupancy, $n_{{\bf k}=0}$. Figure~\ref{fig:fig2}(a) shows that $n_{{\bf k}=0}$ is suppressed with the emergence of a density wave phase reversal in the ground state $|0\rangle$ at large $V_0$'s. Such anticorrelation is present in reverse across selected state in the energy spectrum, particularly for the second excited state $|2\rangle$ [Fig.~\ref{fig:fig2}(b)]~\footnote{The eigenstates listed here obey translation invariance with periodicity ${\cal P}=4$ in $x$-direction and 1 in the $y$-direction and reside in the same quasi-momentum ${\bf q} = 0$.}: the $\pi$-phase shift is absent and $n_{{\bf k}=0}$ is relatively robust, especially at large $V_0/t_h$ [we focus on $V_0=5t_h$ in what follows]. This suggests that fine-tuned perturbations tailored to access the physics of this (and potentially other) excited state(s) are good starting points to tip the balance in favor of enhanced superfluidity. Figures~\ref{fig:fig1}(b) and \ref{fig:fig1}(c) report our main results: The zero-momentum occupancy can be made substantially larger ($\simeq 37\%$) after photoirradiation, at the expense of reduction of the occupation of other momentum modes [Fig.~\ref{fig:fig1}(c)].

Nonetheless, condensation need not be necessarily tied to superfluid behavior, in general~\cite{Fisher1989}. A necessary condition for dissipationless charge transport is given by the charge stiffness (or Drude weight), whose finite value indicates an ideal metal with ballistic transport~\cite{Kohn1964, Scalapino1993, Resta2018}. For the $m$-th eigenstate of $\hat {\cal H}$ with eigenenergy $E_m$,
\begin{equation}
    \frac{D_m^{(\mu)}}{\pi} = \frac{L_\mu^2}{L_xL_y}\left.\frac{\partial^2 E_m}{\partial \Phi_{\mu}^2}\right|_{\Phi_{\mu} = 0} \ ,
    \label{eq:Drude}
\end{equation}
where $\mu = x,y$ and $\Phi_\mu = \int_0^{L_\mu}{\bf A}\cdot d{\bf l}_\mu$ the corresponding flux associated with the $\mu$-direction for a given system size. At finite-temperature $T$~\cite{Castella1995, Zotos1999} or under out-of-equilibrium~\cite{Kaneko2020} conditions, $D = \sum_m p_m D_m$, with  $p_m = e^{-E_m/T}/{\cal Z}$, the Boltzmann weight in the former [$k_B\equiv 1$], and $p_m = |\langle m|\psi(t)\rangle|^2$ in the latter. Figure~\ref{fig:fig1}(d) shows the eigenstate resolved charge stiffness longitudinal to the ladder using a numerical differentiation of the eigenenergies $E_m$ with respect to $\Phi_x$. That the roughly equal distribution of $D_m^{(x)}$ about zero is observed across the spectrum is an indication of the vanishing thermal averaged charge stiffness, as similarly indicated in the one-dimensional \textit{fermionic} Hubbard model~\cite{Carmelo2013, Karrasch2014, Jin2015, Carmelo2018}. 

Under out-of-equilibrium conditions, quantification of charge transport can be accomplished by threading a dynamical flux $\Phi_x(\Delta t)=\delta A_x L_x\Delta t$, associated with an electric field $-\frac{\partial (\Phi_x(t)/L_x)}{\partial t}= \delta A_x$ along the $x$-direction during a time-interval $\Delta t$~\cite{Kaneko2020}. Such a linear-in-time probing vector potential, ${\bf A}(t)=\delta A_x (t - t_1) \hat x$ for $\Delta t = t-t_1>0$, is turned on, in our simulations, much after the center of the photoirradiation pulse, i.e., $t_1\gg t_d$ [see a representation on Fig.~\ref{fig:fig1}(e), top], to quantify if the pulse endowed dissipationless transport properties. This can be obtained in two ways: The first is to recognize that the instantaneous energy $E(t) \equiv \langle \psi(t)|\hat {\cal H}(t)|\psi(t)\rangle$ [Fig.~\ref{fig:fig1}(e)] can, at short-times (or probing fluxes), be expanded as $E(t) = E(t_1+\Delta t) \simeq E(t_1) +\frac{1}{2}\left .\frac{\partial^2 E(t_1+\Delta t)}{\partial \Phi_x^2}\right |_{\Delta t\to 0}\Phi^2_x(\Delta t)$. As such, the energy gain $\Delta E = E(t_1+\Delta t) - E(t_1)$ associated with the probing flux is $\Delta E = \frac{D^{(x)}}{4\pi} \Phi_x^2(\Delta t)$, where we have used the definition of the Drude weight [Eq.~\eqref{eq:Drude}] for energy $E$ and $L_x=2L_y$. 

The inset in Fig.~\ref{fig:fig1}(f) shows the time-dependence of the energy gain during the probe time. Indeed, at short times, a quadratic dependence in $\Phi_x(\Delta t)$ emerges, indicating a finite $D^{(x)}$ and thus dissipationless transport after the pulse. The second estimation of this physical behavior is that $D^{(x)}$ closely matches with an approximant of the out-of-equilibrium Drude weight $\tilde D^{(x)} = \sum_{m = 0}^{199} |\langle m|\psi(t_1)\rangle|^2 D_m^{(x)} = 0.180$, by taking the lowest 200 lowest energy eigenstates of the equilibrium Hamiltonian, which concentrates most of the weight of $|\psi(t)\rangle$ at the time $t=t_1$ of the application of the probing flux --- overlaps $|\langle m|\psi(t_1)\rangle|^2$ are color-mapped in Fig.~\ref{fig:fig1}(d). This is to be contrasted against the equilibrium Drude weight, $D_0^{(x)} = 0.026$, an order-of-magnitude smaller, furnishing evidence for photoinduced ballistic transport for this system size.   

In addition, the energy oscillations in the probing regime with periodicity $2\pi$ in the applied flux [see vertical dotted lines in Fig.~\ref{fig:fig1}(f)] confirm that the fundamental charge in these conditions is $q = 1$, a single hard-core boson. This is a direct generalization to the dynamical setting of the Byers-Yang theorem~\cite{Byers1961, Bloch1970}, establishing that the energy of a system of particles with charge $q$ confined to a closed geometry and subjected to an external flux exhibits periodicity in $\Phi$ with period $\Phi_0 = 2\pi/q$, the flux quantum.

\begin{figure}[t]
    \centering  
    \includegraphics[width=1\columnwidth]{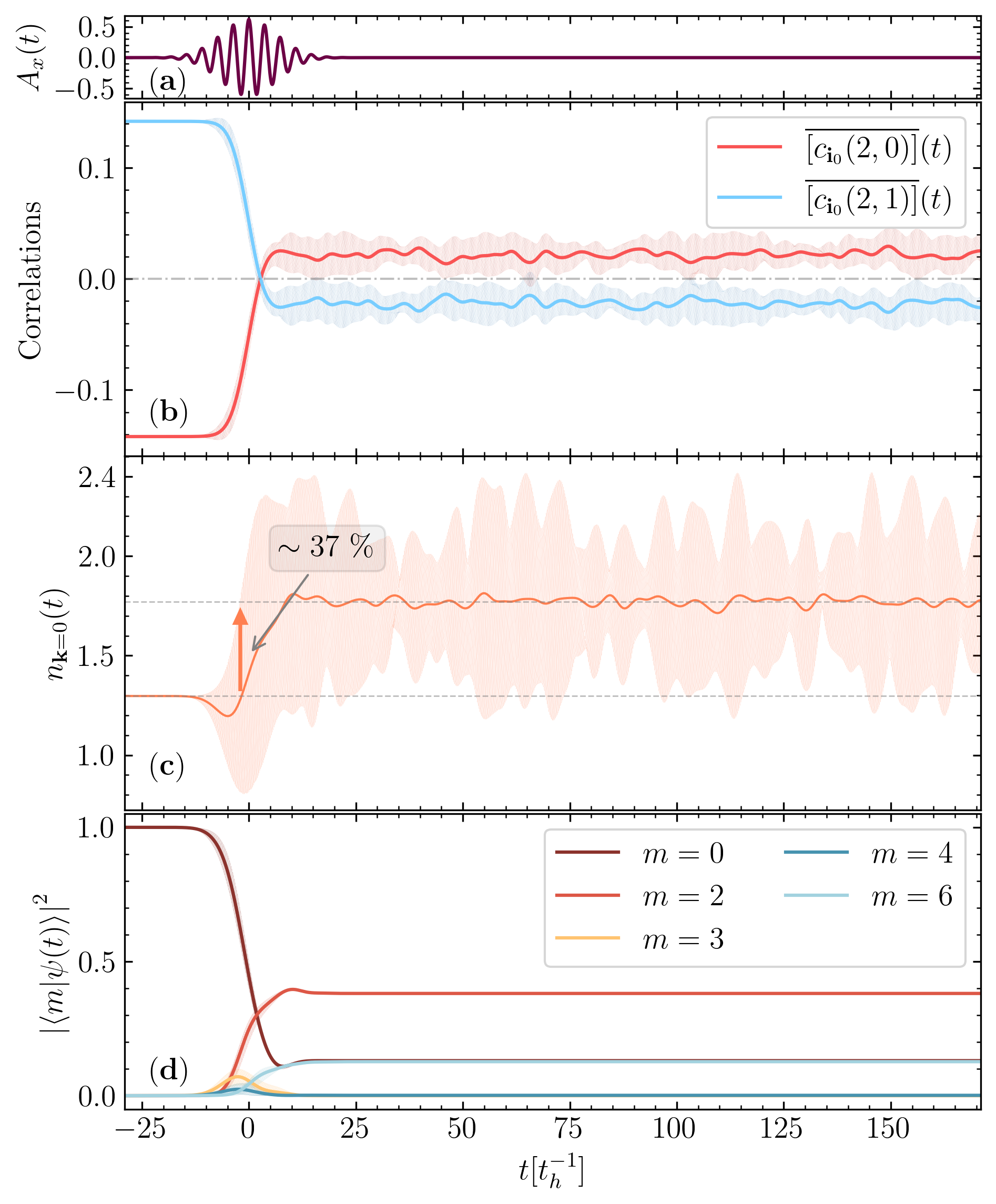}
    \caption{(a) Gaussian pump field with amplitude $A_0=0.62$, frequency $\Omega=\omega_{26}\simeq 1.713\ t_h$, and temporal width $t_d=6.5 t_h^{-1}$ --- the same as in Fig.~\ref{fig:fig1}. Phase-averaged dynamics (solid curves) for a range of phases $\phi_t\in[0,2\pi)$ (shaded regions) are shown in (b)--(d). (b) Dynamics of the correlation functions $c_{{\bf i}_0}(2, 0)$ and $c_{{\bf i}_0}(2, 1)$, in respect to the reference site ${\bf i}_0$ marked in Fig.~\ref{fig:fig1}(a), exhibiting $\pi$-phase shift reversal after the pulse --- the corresponding density dynamics is shown in Appendix \ref{app:dens_dyn}. (c) The zero-momentum occupation $n_{{\bf k}=0}$ exhibits a 37\% enhancement in comparison to the equilibrium. (d) State projection coefficients for selected equilibrium eigenstates $|m\rangle$ (the ones showing significant weight over the dynamics), with state $|2\rangle$ gradually becoming dominant at long times, added by an enhanced overlap with state $|6\rangle$.}
    \label{fig:fig3}
\end{figure} 

\section{Enhanced \tcr{superfluid response}}
The onset of ballistic transport is a necessary, but not sufficient, condition for establishing superfluid behavior. A definitive signature requires a simultaneous enhancement of phase coherence. Figure~\ref{fig:fig1}(f) displays the dynamics of the condensate fraction, $f_0(t)=\lambda_{\rm max}(t)/N$, where $N=L_x\times L_y$ and $\lambda_{\rm max}$ is the largest eigenvalue of the single-particle density matrix $\langle \hat b^{\dagger}_{\bf i} \hat b^{\phantom{\dagger}}_{\bf j}\rangle$. In agreement with the increase in zero-momentum occupation shown in Fig.~\ref{fig:fig1}(b), $f_0(t)$ rises by approximately $34\%$ after photoirradiation, confirming that the buildup of phase coherence indeed accompanies the light-induced enhancement of condensation. This is a direct reflection of the longer-ranged correlations following the pulse --- see Appendix~\ref{app:LR_corr} for an in-depth analysis, including the out-of-equilibrium estimation of the superfluid weight [Appendix \ref{app:D_s_out_of_eq}].
\tcr{It is important to highlight that superfluidity entails long-range order, which can only be obtained with finite-size scalings, elusive for a calculation in out-of-equilibrium. As such, here we refer to an enhanced superfluid \emph{response} on our finite-system size.}

\section{Melting of $\pi$-phase shift} Having confirmed that an appropriate photoirradiation leads to not only a larger occupancy of the zero-momentum mode but also enhanced transport, we now focus on the fate of the charge $\pi$-shift mediated by the stripes. Under the same photoirradiation conditions, later detailed, and apart from the introduction of a probe vector potential, Figure~\ref{fig:fig3} shows the dynamics of $n_{{\bf k}=0}$ closely --- data are reported for a range of the pump's phase shifts $\phi_t$, as well as their average, which aids in smoothening the short-time coherent oscillations. That the photoirradiation can be used to filter pairing in detriment of charge-stripes is apparent by the fact that while $n_{{\bf k}=0}$ dynamically increases [Fig.~\ref{fig:fig3}(c)], the $\pi$-phase shift disappears: $c_{{\bf i}_0}(2,0)$ and $c_{{\bf i}_0}(2,1)$ switch their signs [Fig.~\ref{fig:fig3}(b)] at the time at which the pump [Fig.~\ref{fig:fig3}(a)] has the largest amplitude.

\begin{figure}[t]
    \centering
    \includegraphics[width=1\linewidth]{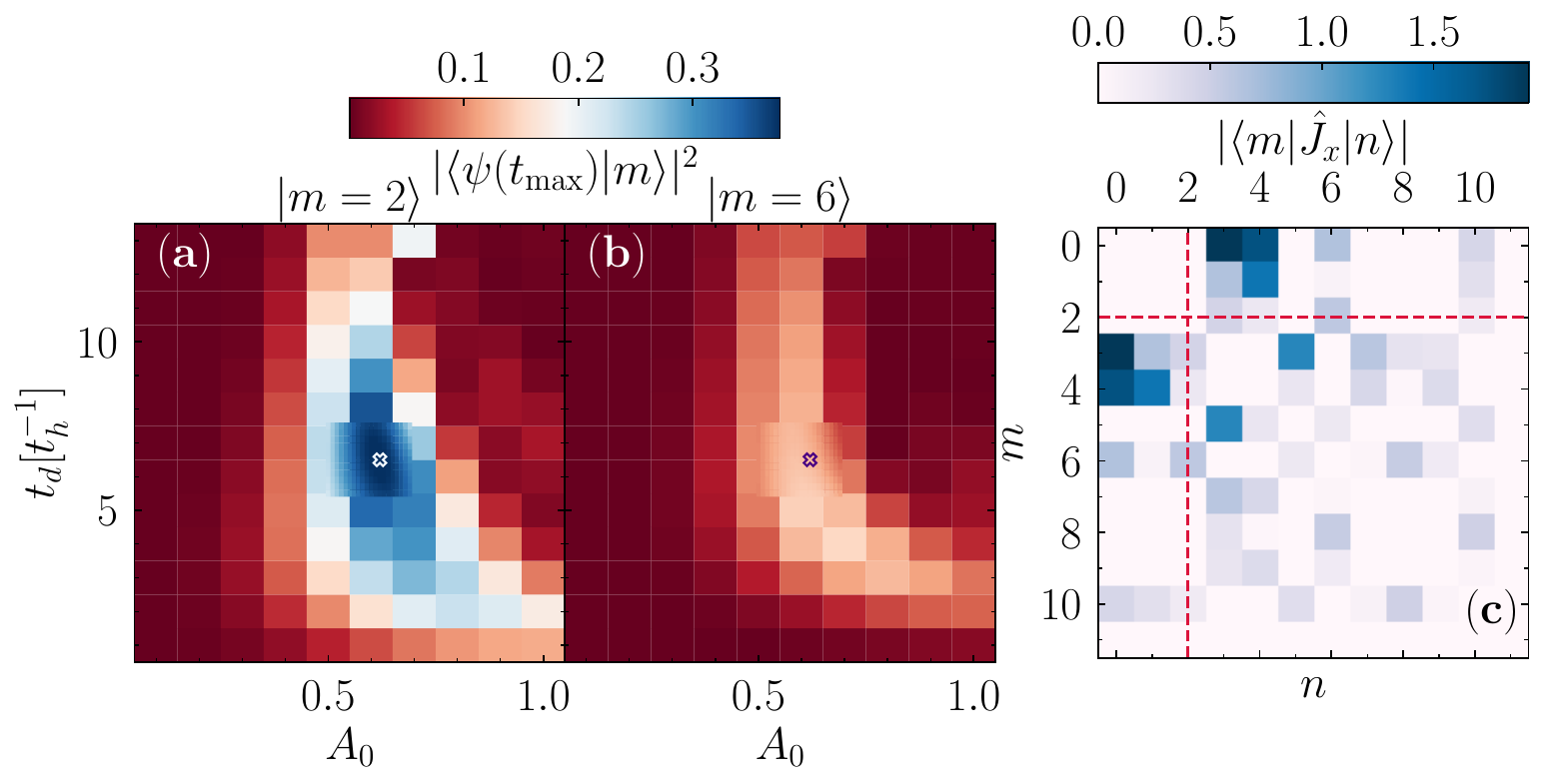}
    \caption{Dependence on the pump width $t_d$ and amplitude $A_0$ of the long-time overlap at $t_{\rm max} = 9t_d$ with selected equilibrium eigenstates, $|m=2\rangle$ (a) and $|m=6\rangle$ (b). A finer mesh resolution is shown in the area with an enhanced overlap $|\langle \psi(t_{\rm max})|2\rangle|^2$ while the cross marker pinpoints the maximum value. Here, the pump frequency is set at $\Omega = \omega_{26} = 1.713\ t_h$. (c) The equilibrium $x$-direction current matrix in the low-lying spectrum; dashed lines highlight matrix elements with the target state $|2\rangle$. Other relevant matrix elements for the perturbative analysis are shown in Appendix \ref{app:mat_elem}.}
    \label{fig:fig4}
\end{figure}

This can be understood in terms of the overlaps $|\langle m|\psi(t)\rangle|^2$ with low-lying equilibrium eigenstates $|m\rangle$ in Fig.~\ref{fig:fig3}(d). At long times, the dynamics stabilize most of the spectral weight in state $|2\rangle$, the target state, as well as $|6\rangle$ with a still relevant weight ($13 \%$) with the initial state, $|0\rangle$. For that, one needs to optimize pump parameters, including the pump's width $t_d$ and amplitude $A_0$, as indicated in Figs.~\ref{fig:fig4}(a) and \ref{fig:fig4}(b), resulting in optimal $A_0 = 0.62$, and $t_d = 6.5 t_h^{-1}$  --- Appendix \ref{app:max_nk0} shows that these pump parameters also maximize the long-time average of $n_{{\bf k}=0}$. However, additional setting of the pump's frequency is required to achieve the most accurate targeting of the excited state. Previous investigations in various models~\cite{Shao2019, Shao2021, Yuxi2023} established that a resonant photoirradiation, $\Omega = E_{\rm target}-E_0$, could accomplish a direct targeting of relevant excited states with eigenenergies $E_{\rm target}$. At the crux of such a process is the fact that first-order time-dependent perturbation theory stemming from the pulse application (details in Appendix \ref{app:t_pert}) establishes that the $x$-direction current operator, $\hat J_x = it_h\sum_{\bf i}(\hat b_{\bf i+\hat x}^\dagger\hat b_{\bf i}^{\phantom{\dagger}} -  b_{\bf i}^\dagger\hat b_{\bf i+\hat x}^{\phantom{\dagger}})$, connects these states with a robust $|\langle m_{\rm target}|\hat J_{x}|0\rangle|$. Unfortunately, our investigation shows that, in our case, the matrix elements of the current operator between $|0\rangle$ and $|2\rangle$ are small [Fig.~\ref{fig:fig4}(c)].

\section{The non-linear optical coupling} 
\begin{figure}[t!]
    \centering
    \includegraphics[width=1\linewidth]{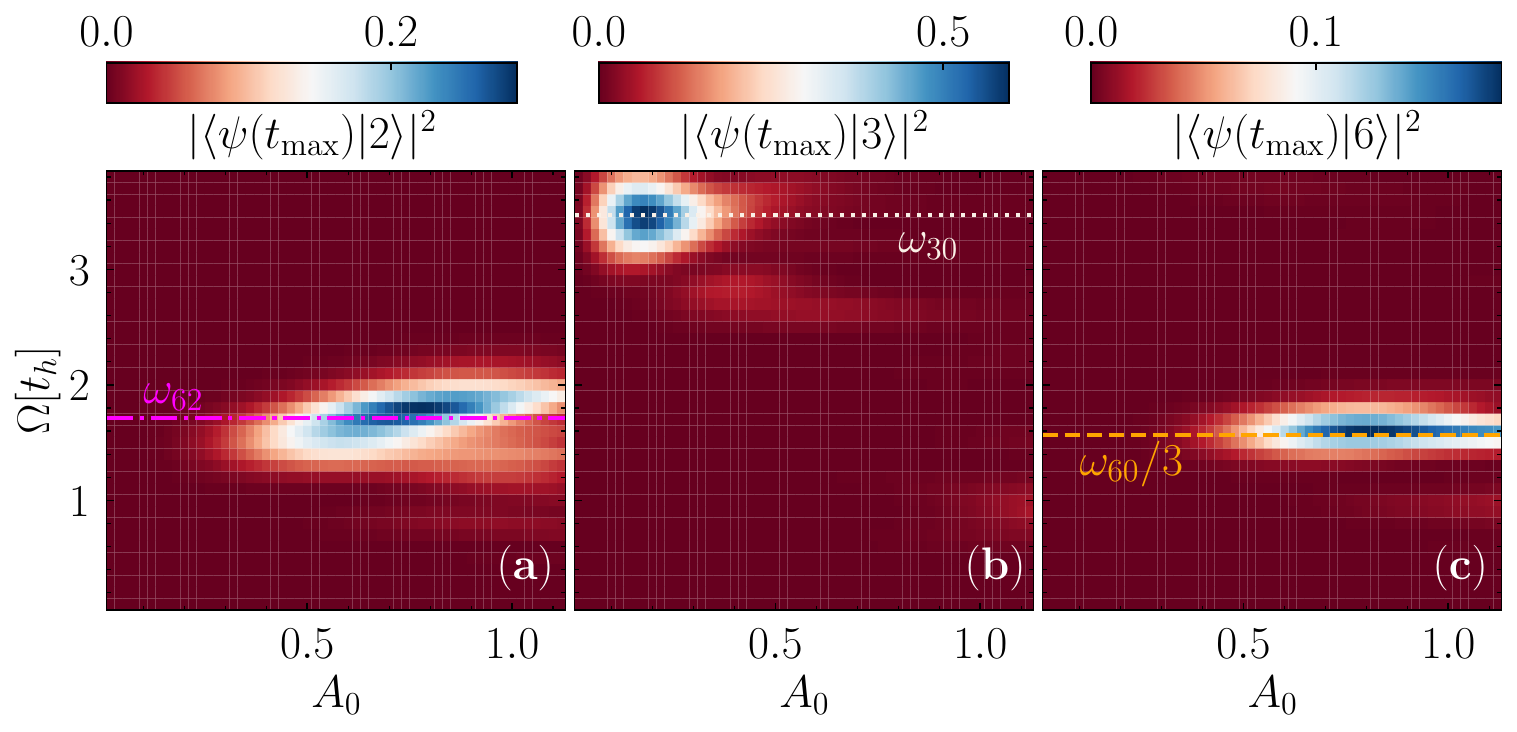}
    \caption{Dependence on the pump frequency $\Omega$ and amplitude $A_0$ of the long-time overlap at $t_{\rm max} = 9t_d = 27 t_h^{-1}$ with selected equilibrium eigenstates, $|m=2\rangle$ (a), $|m=3\rangle$ (b), and $|m=6\rangle$ (c). Horizontal lines indicate relevant gaps in the equilibrium spectrum; note the different ranges of the color bars, which are set to enhance visualization.}
    \label{fig:fig5}
\end{figure}

That resonant targeting cannot couple to the target state does not impede higher-order, and thus nonlinear, optical excitation from taking place. This is particularly possible because intermediate equilibrium states can have a relatively large current matrix element, e.g., $|\langle 0|\hat J_x|6\rangle| \simeq 0.66$ and $|\langle 6|\hat J_x|2\rangle| \simeq 0.58$. Indeed, in second-order time-dependent perturbation theory, the time-dependent expansion coefficients in the interaction picture entering in $|\psi_{\rm I}(t)\rangle = \sum_{m}c_m(t)|m\rangle$, for initial state $|0\rangle$, final state $|2\rangle$, intermediate state $|m_1\rangle$, at infinite time, are (see Appendix \ref{app:t_pert})
\begin{align}
c_{m=2}^{(2)}(\infty) 
= \sum_{m_1} \sum_{n_1, n_2} 
&\, \hat{V}_{2, m_1}^{(n_2)} \hat{V}_{m_1, 0}^{(n_1)} 
\cdot \frac{1}{E_{m_1} - E_0 - n_1 \Omega - i\epsilon} \notag \\
&\times \delta\left(E_2 - E_0 - (n_1 + n_2) \Omega\right)\ ,
\label{eq:sec_ord}
\end{align}
where $\hat V_{m,m^\prime}^{(n)}$ are the matrix elements of the $ n$-th Fourier component of the perturbation in the unperturbed eigenstates $|m\rangle$ of $\hat {\cal H}$, and the delta function enforces total energy conservation via the absorption or emission of $n_1+n_2$ photons. Now, considering the intermediate state $|m_1=6\rangle$ has a significant contribution, a close resonance occurs for $E_6 \approx E_0+n_1\Omega$, when the denominator in Eq.~\eqref{eq:sec_ord} is small. Meanwhile, the energy conservation condition yields $E_2 = E_0 + (n_1 + n_2)\Omega$. Figure~\ref{fig:fig5}(a) shows that at pump frequencies $\Omega=\omega_{62}\equiv E_6-E_2$, a large overlap with the target state $|2\rangle$ is obtained at long times. Using the two previous energy conditions, $\omega_{62}=-n_2\Omega$, indicating that $n_2=-1$ (a photon emission in between the intermediate and final states), explains the observations. As a result, the resonance condition leads to $E_6\approx E_0+n_1\omega_{62}$, therefore, $\omega_{60}/\omega_{62} \approx n_1$. The equilibrium eigenspectrum yields $\omega_{60}/\omega_{62}\simeq 2.75$, indicating that the resonant excitation process from the ground-state to the intermediate one likely involves the absorption of three quanta of energy $\Omega = \omega_{62}$. Resonance conditions with other intermediate states are explored in Appendix \ref{app:other_interm}.

Lastly, to confirm this perturbation picture, the perturbation matrix elements explain an enlarged overlap at long times with other states as well, but at \textit{first-order}: a considerable overlap with state $|6\rangle$ [Fig.~\ref{fig:fig5}(c)] is obtained at the resonance $\omega_{60}/\Omega_{\rm pump} = n_1 = 4.69/1.57\simeq 3$,  via a three quanta absorption. In turn, the most prominent matrix element of $\hat J_x$ in the low-lying spectrum, $|\langle 3|\hat J_x|0\rangle|\simeq1.93$, can be dynamically accessed in first-order with pump frequencies $\Omega_{\rm pump} \simeq \omega_{30}$ --- Figure~\ref{fig:fig5}(b) precisely confirms this scenario. Analysis in terms of the symmetries of the operators involved in this perturbative picture further corroborates these results, and, together with the corresponding parity of the low-lying states targeted in the pump process, is detailed in Appendix~\ref{app:symmetries}.

\section{Summary and outlook}  We showed that fine-tuned photoirradiation, effectively promoting a non-linear optical excitation towards a target state, forms a pathway for dynamically tailoring a system to exhibit desired properties. This was exemplified in the case of the competing interplay between pairing and density-wave with phase reversal in the presence of stripes; however, the approach is likely applicable in other scenarios. Key is the identification of the equilibrium properties of the low-lying eigenspectrum, and the design of a perturbation that, in sufficiently low order, can access them. Notably, that generically melting stripes leads to transient superconductivity is a result under intense debate in experiments~\cite{Zhang2024}. Our results affirmatively point to this connection in a specific model. Confirming this for the fermionic Hubbard and $t$-$J$ models relevant to the cuprates, and larger system sizes,  requires numerics currently beyond the capabilities of exact diagonalization and Krylov methods we employ. This is currently only possible with other techniques that approximate the dynamics, such as the time-dependent variational principle in the language of matrix product states~\cite{Haegeman2011, Haegeman2016, Kloss2018, Li2024} --- we leave this for future investigation.

\begin{acknowledgments}R.M.~acknowledges support from the T$_c$SUH Welch Professorship Award. R.T.S. is supported by the grant DOE DE-SC0014671 funded by the U.S. Department of Energy, Office of Science. Numerical simulations were performed with resources provided by the Research Computing Data Core at the University of Houston. This work also used TAMU ACES at Texas A\&M HPRC through allocation PHY240046 from the Advanced Cyberinfrastructure Coordination Ecosystem: Services \& Support (ACCESS) program, which is supported by U.S. National Science Foundation grants 2138259, 2138286, 2138307, 2137603, and 2138296. The data that support the findings of this article are openly available~\cite{mondaini2025_zenodo}.
\end{acknowledgments}


\appendix
\section{Time-dependent Perturbation Theory for the photoirradiation}
\label{app:t_pert}
The main text explains how to enhance zero-momentum occupancy and melt the charge $\pi$-phase shift by targeting eigenstates of the low-lying equilibrium energy spectrum. The manner by which this targeting is achieved is through the tuning of the pulse properties, in particular its frequency, such that at long times, the weight of the target state in the time-evolving wavefunction $|\psi(t)\rangle$ is maximized. We now explain how this arises as a consequence of time-dependent perturbation theory applied to the pulsed driving --- Ref.~\cite{Shirakawa2020} describes an equivalent analysis in another lattice model.

The effect of the pulse is to modify the equilibrium Hamiltonian $\hat {\cal H}$ [Eq.~\eqref{eq:model}] with a time-dependent perturbation $\hat{V}(t)$,
\begin{equation}
    \hat H(t) = \hat {\cal H} + \hat{V}(t)\ .
    \label{eq:perturbation}
\end{equation} 
Considering that the eigenstates $|m\rangle$ of $\hat {\cal H}$  define a basis, the time-evolved state at arbitrary times takes the form
\begin{equation}
    |\psi(t)\rangle = \sum_m c_m(t) e^{-i E_m t} |m\rangle\ ,
\end{equation}
where $c_m(t)$ are the time-dependent expansion coefficients in the Schrödinger picture, related to the interaction picture amplitudes by $\langle m|\psi_{\rm I}(t)\rangle=e^{iE_mt}c_m(t)$. In this latter picture, the state vector evolves according to,
\begin{equation}
    i \frac{d}{dt} |\psi_{\rm I}(t)\rangle = \hat{V}_{\rm I}(t) |\psi_{\rm I}(t)\rangle,
\end{equation}
with $\hat{V}_{\rm I}(t) = e^{i \hat{\cal H} t} \hat{V}(t) e^{-i\hat{\cal H} t}$ and $|\psi_{\rm I}(t)\rangle = e^{i \hat{\cal H} t} |\psi(t)\rangle$. In particular, recalling that the ground-state $|0\rangle$ coincides with the initial state of the system, $|\psi(t\to-\infty)\rangle$, the interaction picture state can be expanded in a Dyson series~\cite{Sakurai1994}:
\begin{align}
    |\psi_{\rm I}(t)\rangle &= \mathcal{T} \exp\left(-i \int_{-\infty}^{t} dt' \hat{V}_{\rm I}(t') \right) |0\rangle\notag \\
    &= |0\rangle + |\psi_{\rm I}^{(1)}(t)\rangle + |\psi_{\rm I}^{(2)}(t)\rangle + \ldots\ ,
\end{align}
where \( \mathcal{T} \) is the time-ordering operator. To first order in the perturbation, one obtains
\begin{equation}
    |\psi_{\rm I}^{(1)}(t)\rangle = -i \int_{-\infty}^{t} dt_1 \, \hat{V}_{\rm I}(t_1) |0\rangle\ , 
\end{equation}
whereas the second-order contribution yields,
\begin{equation}
    |\psi_{\rm I}^{(2)}(t)\rangle = (-i)^2 \int_{-\infty}^{t} dt_2 \int_{-\infty}^{t_2} dt_1 \, \hat{V}_{\rm I}(t_2) \hat{V}_{\rm I}(t_1) |0\rangle\ .
\end{equation}

Focusing on the first order, the amplitude to find the system in the unperturbed state $|m\rangle$, $\langle m|\psi_{\rm I}^{(1)}(t)\rangle\equiv c_m^{(1)}(t)$, becomes
\begin{equation}
    c_m^{(1)}(t) = -i \int_{-\infty}^{t} dt_1 \, e^{i E_m t_1} \langle m|\hat{V}(t_1)|0\rangle e^{-i E_0 t_1}\ .
    \label{eq:first_order}
\end{equation}
In turn, the second-order amplitude reads,
\begin{align}
    c_m^{(2)}(t) = (-i)^2 \sum_{m^\prime} &\int_{-\infty}^{t} dt_2 \int_{-\infty}^{t_2} dt_1 \, \notag \\
    &\times e^{i E_m t_2} \langle m|\hat{V}(t_2)|m^\prime\rangle e^{-i E_{m^\prime} t_2} \notag \\
    &\times e^{i E_{m^\prime} t_1} \langle m^\prime|\hat{V}(t_1)|0\rangle e^{-i E_0 t_1}\ ,
\end{align}
where the index $m^\prime$ runs over all intermediate eigenstates of the unperturbed Hamiltonian. 

If the perturbation is rewritten in terms of its Fourier series,
\begin{equation}
 \hat{V}(t) = \sum_{n=-\infty}^\infty \hat V_ne^{-in\Omega t}, 
 \label{eq:disc_FT}
\end{equation}
then, in the infinite-time limit, the first-order transition amplitude [Eq.\eqref{eq:first_order}], after the time integration~\footnote{Using the identity $\int_{-\infty}^\infty dt e^{i\omega t}=2\pi\delta(\omega)$}, becomes
\begin{equation}
    c_m^{(1)}(\infty) = -2\pi i \sum_{n} 
    \langle m|\hat{V}_n|0\rangle \, \delta(E_m - E_0 - n\Omega)\ ,
    \label{eq:first_order_amplitude}
\end{equation}
which reflects energy conservation for the absorption and emission of $n$ photons of frequency $\Omega$.

Similarly, the second-order transition amplitude becomes
\begin{align}
    c_m^{(2)}(\infty) = - \sum_{n_1, n_2} \sum_{m^\prime}
    &\frac{\langle m|\hat{V}_{n_2}|m^\prime\rangle \langle m^\prime|\hat{V}_{n_1}|0\rangle} 
    {E_{m^\prime} - E_0 - n_1 \Omega - i \epsilon} \,\notag \\ 
    &\times \delta(E_m - E_0 - (n_1 + n_2)\Omega)\ ,
\end{align}
where the sum runs over all intermediate equilibrium states \( |m^\prime\rangle \), and the delta function enforces energy conservation for the net exchange of \( n_1 + n_2 \) photons.

Now that we have established the perturbative coefficients, we are in a position to describe their complete functional form. First, we recall the structure of $\hat V(t)$ in Eq.~\eqref{eq:perturbation}, which arises from the hopping modification of the kinetic part of the Hamiltonian [$\hat H(t) = \hat K(t)+\hat V(t)$]:
\begin{align}
\hat K(t) &= -t_h \sum_{\langle {\bf i},{\bf j} \rangle} \Big[ 
e^{iA_{{\bf ij}}(t)} \hat{b}^\dagger_{\bf i} \hat{b}_{\bf j}^{\phantom{\dagger}} + e^{-iA_{{\bf ij}}(t)}\hat{b}^\dagger_{\bf j} \hat{b}_{\bf i}^{\phantom{\dagger}}\Big]\notag\\
&=-t_h\sum_{\langle {\bf i},{\bf j} \rangle} \Big[\cos A_{{\bf ij}}(t)\left(\hat{b}^\dagger_{\bf i} \hat{b}_{\bf j}^{\phantom{\dagger}}+\hat{b}^\dagger_{\bf j}\hat{b}_{\bf i}^{\phantom{\dagger}}\right)\notag \\
&\quad\quad\quad\quad+i\sin A_{{\bf ij}}(t)\left(\hat{b}^\dagger_{\bf i} \hat{b}_{\bf j}^{\phantom{\dagger}}-\hat{b}^\dagger_{\bf j}\hat{b}_{\bf i}^{\phantom{\dagger}}\right)\Big]\ ,
\label{eq:second_order_amplitude}
\end{align}
where $A_{{\bf ij}}(t) = {\bf A}(t)\cdot({\bf i}-{\bf j})$. As a result, the time-dependent perturbation in Eq.~\eqref{eq:perturbation} yields,
\begin{align}
    \hat V(t)=-t_h\sum_{\langle {\bf i},{\bf j} \rangle} \Big[\left(\cos A_{{\bf ij}}(t)-1\right)&\left(\hat{b}^\dagger_{\bf i} \hat{b}_{\bf j}^{\phantom{\dagger}}+\hat{b}^\dagger_{\bf j}\hat{b}_{\bf i}^{\phantom{\dagger}}\right)\notag\\
    +i\sin A_{{\bf ij}}(t)&\left(\hat{b}^\dagger_{\bf i} \hat{b}_{\bf j}^{\phantom{\dagger}}-\hat{b}^\dagger_{\bf j}\hat{b}_{\bf i}^{\phantom{\dagger}}\right)\Big]\ ,
\end{align}
which is a sum of a time-modulated kinetic-energy and current-like terms for the direction in which the vector potential is applied. For instance, for our pulse polarized along the $x$-direction, ${\bf A}(t) = A(t)\hat {\bf x}$, $A_{{\bf ij}}(t) = A(t)$ if ${\bf j}={\bf i}+ \hat {\bf x}$, and $A_{{\bf ij}}(t) = 0$ if  ${\bf j}={\bf i}+ \hat {\bf y}$. As a result, the time-dependent perturbation can be written as $\hat V(t) = \hat K_x(t)+\hat J_x(t)$, where 
\begin{equation}
    \hat K_x(t) \equiv -t_h\left(\cos A(t)-1\right)\sum_{\bf i} \left(\hat{b}^\dagger_{\bf i} \hat{b}_{{\bf i}+\hat {\bf x}}^{\phantom{\dagger}}+\hat{b}^\dagger_{{\bf i}+\hat {\bf x}}\hat{b}_{\bf i}^{\phantom{\dagger}}\right)
\end{equation}
and,
\begin{equation}
  \hat J_x(t) \equiv -it_h\sin A(t)\sum_{\bf i} \left(\hat{b}^\dagger_{\bf i} \hat{b}_{{\bf i}+\hat {\bf x}}^{\phantom{\dagger}}-\hat{b}^\dagger_{{\bf i}+\hat {\bf x}}\hat{b}_{\bf i}^{\phantom{\dagger}}\right)\ .
\end{equation}
Now there are two choices to achieve analytical expressions for the transition amplitudes: The first is to assume the perturbation is perfectly time-periodic, using the discrete Fourier transform [Eq.~\eqref{eq:disc_FT}] to fully describe the Floquet-like driving of the system, i.e., by taking the pulse width $t_d\to\infty$, i.e., $A(t) \simeq A_0\cos(\Omega t)$ and the second is to taking into account the Gaussian envelope that modulates the sinusoidal pulse function. In the former fully periodic case, the discrete Fourier coefficients in Eq.~\ref{eq:disc_FT}, read
\begin{equation}
    \hat V_n = \frac{1}{T}\int_0^T dt \ \hat V(t)e^{in\Omega t},\quad \text{with,}\quad T=2\pi/\Omega.
\end{equation}
Equivalently, one can deduce these coefficients by making use of the Jacobi-Anger expansions~\cite{Abramowitz1964}, for the relevant terms in $\hat K_x(t)$ and $\hat J_x(t)$:
\begin{align}
    \cos(A_0\cos(\Omega t))-1 &= \mathcal{J}_0(A_0)-1 \notag \\ &\quad+ 2\sum_{n=1}^\infty\mathcal{J}_{2n}(A_0)\cos(2n\Omega t) \notag\\
    &=\sum_{m=-\infty}^\infty a_m e^{-i m\Omega t}\ ,
\end{align}
and 
\begin{align}
    \sin(A_0\cos(\Omega t)) &= 2\sum_{n=0}^\infty \mathcal{J}_{2n+1}(A_0)\sin((2n+1)\Omega t) \notag \\
    &=\sum_{m=-\infty}^\infty b_m e^{-i m\Omega t}\ ,
\end{align}
where $\mathcal{J}_n(x)$ are the Bessel functions of the first kind of integer order $n$. After expanding $\cos(2n\Omega t)$ and $\sin((2n+1)\Omega t)$ to their exponential form, it leads to the identification $a_0 = {\cal J}_0(A_0)-1$, $a_{\pm 2n}={\cal J}_{2n}(A_0)$, $a_{\pm(2n+1)}=0$ and $b_{\pm (2n+1)}=\mp i{\cal J}_{2n+1}(A_0)$, $b_{2n}=0$. This fully defines thus the terms entering in the transition amplitudes in Eqs.~\eqref{eq:first_order_amplitude} and \eqref{eq:second_order_amplitude} as,
\begin{align}
    \hat V_n = &-t_h a_n \sum_{\bf i} \left(\hat{b}^\dagger_{\bf i} \hat{b}_{{\bf i}+\hat {\bf x}}^{\phantom{\dagger}}+\hat{b}^\dagger_{{\bf i}+\hat {\bf x}}\hat{b}_{\bf i}^{\phantom{\dagger}}\right) \notag \\
    &-it_h b_n\sum_{\bf i} \left(\hat{b}^\dagger_{\bf i} \hat{b}_{{\bf i}+\hat {\bf x}}^{\phantom{\dagger}}-\hat{b}^\dagger_{{\bf i}+\hat {\bf x}}\hat{b}_{\bf i}^{\phantom{\dagger}}\right)\ .
\end{align}

In the second case, where $A(t)=A_0e^{-t^2/2t_d^2}\cos(\Omega t)$, instead of using the Fourier expansion~\footnote{Which for this case should be instead a continuous Fourier transform unlike the discrete case of Eq.~\eqref{eq:disc_FT}.}, insight can be gained by taking the limit of small pulse amplitudes, $A_0\ll 1$, such that $\cos A(t) \approx 1 - \frac{1}{2}A^2(t)$ and $\sin A(t) \approx A(t)$, leading to
\begin{align}
    \hat K_x(t) &\approx \frac{t_h}{2}A^2(t)\sum_{\bf i} \left(\hat{b}^\dagger_{\bf i} \hat{b}_{{\bf i}+\hat {\bf x}}^{\phantom{\dagger}}+\hat{b}^\dagger_{{\bf i}+\hat {\bf x}}\hat{b}_{\bf i}^{\phantom{\dagger}}\right) \notag \\
    &=\frac{t_h}{2}A_0^2e^{-t^2/t_d^2}[1+\cos(2\Omega t)]\left(\hat{b}^\dagger_{\bf i} \hat{b}_{{\bf i}+\hat {\bf x}}^{\phantom{\dagger}}+\hat{b}^\dagger_{{\bf i}+\hat {\bf x}}\hat{b}_{\bf i}^{\phantom{\dagger}}\right) \notag \\
    &\equiv A^2(t)\hat K_x^{(2)}\ ,
\end{align}
and 
\begin{align}
    \hat J_x(t) &\approx -it_h A(t)\sum_{\bf i} \left(\hat{b}^\dagger_{\bf i} \hat{b}_{{\bf i}+\hat {\bf x}}^{\phantom{\dagger}}-\hat{b}^\dagger_{{\bf i}+\hat {\bf x}}\hat{b}_{\bf i}^{\phantom{\dagger}}\right)\notag \\
    & = -it_h A_0e^{-t^2/2t_d^2}\cos(\Omega t)\sum_{\bf i}\left(\hat{b}^\dagger_{\bf i} \hat{b}_{{\bf i}+\hat {\bf x}}^{\phantom{\dagger}}-\hat{b}^\dagger_{{\bf i}+\hat {\bf x}}\hat{b}_{\bf i}^{\phantom{\dagger}}\right)\notag\\
    & \equiv A(t)\hat J_x^{(2)}\ .
\end{align}
As a result, the first-order transition amplitude [Eq.~\eqref{eq:first_order} for $t\to\infty$] becomes:
\begin{widetext}
\begin{align}
c_{m}^{(1)}(\infty) &\simeq -i \int_{-\infty}^{\infty} dt\, \Big[
A(t) \langle m|\hat{J}_x^{(2)}|0\rangle 
+ A^2(t) \langle m|\hat{K}_x^{(2)}|0\rangle
\Big] e^{i \omega_{m0} t} \notag \\
&= -i t_d \Bigg\{ 
\frac{1}{2}A_0 \sqrt{2\pi} 
\left[ 
e^{-\frac{t_d^2}{2} (\omega_{m0} - \Omega)^2} 
+ e^{-\frac{t_d^2}{2} (\omega_{m0} + \Omega)^2} 
\right] \langle m | \hat{J}_x^{(2)} | 0 \rangle \notag \\
&\quad + A_0^2 \frac{\sqrt{\pi}}{2} 
\left[ 
e^{-\frac{t_d^2}{4} \omega_{m0}^2} 
+ \frac{1}{2} \left( 
e^{-\frac{t_d^2}{4} (\omega_{m0} - 2\Omega)^2} 
+ e^{-\frac{t_d^2}{4} (\omega_{m0} + 2\Omega)^2} 
\right) 
\right] \langle m | \hat{K}_x^{(2)} | 0 \rangle 
\Bigg\}\ ,
\label{eq:first_order_via_pulse}
\end{align}
\end{widetext}
where we use the main text notation $\omega_{m0}\equiv E_m-E_0$. The resulting transition probability is then governed by the spectral content of $A(t)$ and $A^2(t)$, which are sharply peaked around $\Omega$ and $2\Omega$, respectively.

Likewise, in the second-order expansion under the small-$A_0$ approximation, the transition amplitude becomes
\begin{widetext}
\begin{align}
    c_{m}^{(2)}(\infty) \simeq & -\sum_{m^\prime}\int_{-\infty}^{\infty}dt_2\int_{-\infty}^{t_2}dt_1\,
e^{i\omega_{mm^\prime}t_2}e^{i\omega_{m^\prime0}t_1} \notag \\
&\times \Big[
A(t_2)A(t_1)\langle m|\hat J_x^{(2)}|m^\prime\rangle\langle m^\prime|\hat J_x^{(2)}|0\rangle + A(t_2)A^2(t_1)\langle m|\hat J_x^{(2)}|m^\prime\rangle\langle m^\prime|\hat K_x^{(2)}|0\rangle \notag \\
&\quad+ A^2(t_2)A(t_1)\langle m|\hat K_x^{(2)}|m^\prime\rangle\langle m^\prime|\hat J_x^{(2)}|0\rangle + A^2(t_2)A^2(t_1)\langle m|\hat K_x^{(2)}|m^\prime\rangle\langle m^\prime|\hat K_x^{(2)}|0\rangle
\Big] \notag \\
=& -\sum_{m^\prime} \Big[
\langle m| \hat{J}_x^{(2)} | m^\prime \rangle \langle m^\prime | \hat{J}_x^{(2)} | 0 \rangle \, \mathcal{I}_{JJ}^{(m^\prime)} + \langle m | \hat{K}_x^{(2)} | m^\prime \rangle \langle m^\prime | \hat{K}_x^{(2)} | 0 \rangle \, \mathcal{I}_{KK}^{(m^\prime)} \notag \\
&\quad+ \left( \langle m | \hat{J}_x^{(2)} | m^\prime \rangle \langle m^\prime | \hat{K}_x^{(2)} | 0 \rangle + \text{H.c.} \right)
\, \mathcal{I}_{JK}^{(m^\prime)} \Big]\ ,
\end{align}
\end{widetext}
where the time integrals are weighted convolutions of Gaussian wavepackets centered at those harmonics, and are approximated by
\begin{widetext}
\begin{align}
\mathcal{I}_{JJ}^{(m')} &= \int_{-\infty}^{\infty} dt_2 \int_{-\infty}^{t_2} dt_1 \,
e^{- \frac{t_1^2 + t_2^2}{2t_d^2}} \cos(\Omega t_1) \cos(\Omega t_2)
\, e^{i\omega_{m'0}t_1} e^{i\omega_{mm'}t_2}  \notag \\
&\approx \frac{\pi t_d^2}{8}\left[
e^{-\frac{t_d^2}{2}\left[(\omega_{m'0}+\Omega)^2+(\omega_{mm'}+\Omega)^2\right]}\right.
+e^{-\frac{t_d^2}{2}\left[(\omega_{m'0}-\Omega)^2+(\omega_{mm'}+\Omega)^2\right]}\notag \\
&\quad+e^{-\frac{t_d^2}{2}\left[(\omega_{m'0}+\Omega)^2+(\omega_{mm'}-\Omega)^2\right]}\left.+e^{-\frac{t_d^2}{2}\left[(\omega_{m'0}-\Omega)^2+(\omega_{mm'}-\Omega)^2\right]}
\right]\ ,
\label{eq:second_order_via_pulse}
\end{align}
where we have assumed in the integration that $t_d\Omega \gg 1$. Likewise,
\begin{align}
\mathcal{I}_{KK}^{(m')} &= \int_{-\infty}^{\infty} dt_2 \int_{-\infty}^{t_2} dt_1 \,
e^{- \frac{t_1^2 + t_2^2}{t_d^2}} [1 + \cos(2\Omega t_1)][1 + \cos(2\Omega t_2)] 
\, e^{i\omega_{m'0} t_1} e^{i\omega_{mm'} t_2} \notag \\
&\approx \frac{\pi t_d^2}{2} e^{-\frac{t_d^2}{4}(\omega_{m'0}^2+\omega_{mm'}^2)} +\frac{\pi t_d^2}{4}\left[e^{-\frac{t_d^2}{4}[(\omega_{m'0}+2\Omega)^2+\omega_{mm'}^2]}+e^{-\frac{t_d^2}{4}[(\omega_{m'0}-2\Omega)^2+\omega_{mm'}^2]}\right]\notag\\
&+\frac{\pi t_d^2}{4}\left[e^{-\frac{t_d^2}{4}[(\omega_{mm'}+2\Omega)^2+\omega_{m'0}^2]}+e^{-\frac{t_d^2}{4}[(\omega_{mm'}-2\Omega)^2+\omega_{m'0}^2]}\right]\notag\\
&+\frac{\pi t_d^2}{8}\biggl[
e^{-\frac{t_d^2}{4}[(\omega_{m'0}+2\Omega)^2+(\omega_{mm'}+2\Omega)^2]}+e^{-\frac{t_d^2}{4}[(\omega_{m'0}+2\Omega)^2+(\omega_{mm'}-2\Omega)^2]}\notag\\
&\quad\quad+e^{-\frac{t_d^2}{4}[(\omega_{m'0}-2\Omega)^2+(\omega_{mm'}+2\Omega)^2]}+e^{-\frac{t_d^2}{4}[(\omega_{m'0}-2\Omega)^2+(\omega_{mm'}-2\Omega)^2]}
\biggr]\ ,
\end{align}
and
\begin{align}
\mathcal{I}_{JK}^{(m')} &= \int_{-\infty}^{\infty} dt_2 \int_{-\infty}^{t_2} dt_1 \,
e^{- \frac{t_2^2}{2t_d^2}} \cos(\Omega t_2) \,
e^{- \frac{t_1^2}{t_d^2}} [1 + \cos(2\Omega t_1)] \,
e^{i\omega_{m'0} t_1} e^{i\omega_{mm'} t_2} \notag \\
&\approx \frac{\pi t_d}{2} \, e^{-\frac{t_d^2}{4} \omega_{m'0}^2}
\left[
e^{-\frac{t_d^2}{2} (\omega_{mm'} - \Omega)^2}
+ 
e^{-\frac{t_d^2}{2} (\omega_{mm'} + \Omega)^2}
\right] \notag \\
&\quad + \frac{\pi t_d}{4} 
\left[
e^{-\frac{t_d^2}{4} (\omega_{m'0} - 2\Omega)^2}
+ 
e^{-\frac{t_d^2}{4} (\omega_{m'0} + 2\Omega)^2}
\right]
\left[
e^{-\frac{t_d^2}{2} (\omega_{mm'} - \Omega)^2}
+ 
e^{-\frac{t_d^2}{2} (\omega_{mm'} + \Omega)^2}
\right]\ ,
\end{align}
\end{widetext}
denoting resonant processes at different frequencies. For instance, the $JJ$ case, leads to one-photon resonances at $\omega_{m'0}\approx \pm \Omega$ and $\omega_{mm'}\approx \pm \Omega$. The Gaussian envelope sharply restricts these resonances, ensuring that only states close to these conditions contribute significantly. The $KK$ case allows for two-photon processes or virtual zero-energy components, with resonances occurring at $\omega_{m'0},\omega_{mm'}\approx 0, \pm 2\Omega$, and the mixed case $JK$ couples one-photon and two-photon contributions, describing interference between different harmonics.

\section{Relevant Matrix elements}
\label{app:mat_elem}
\begin{figure}[!t]
    \centering
    \includegraphics[width=1\linewidth]{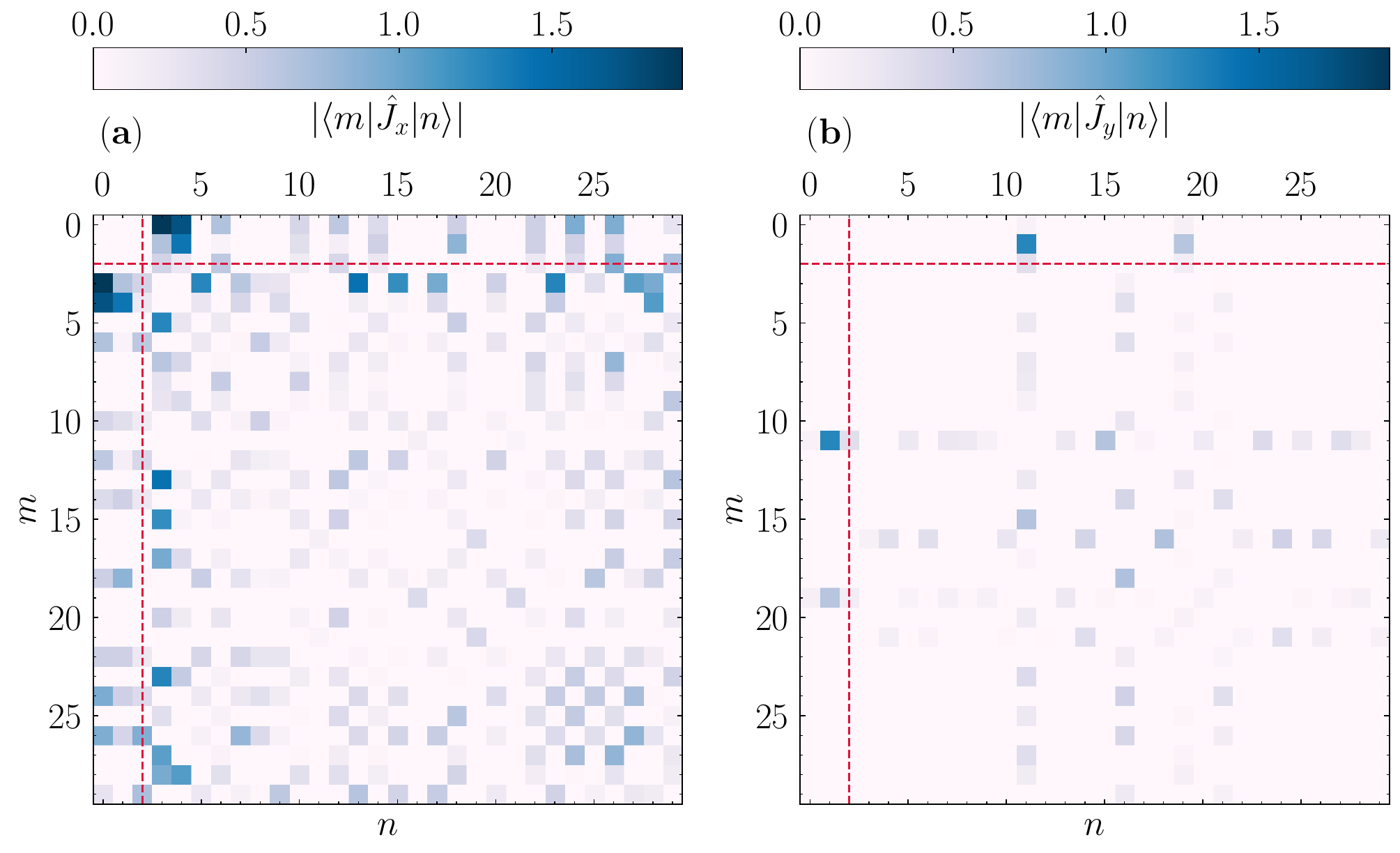}
    \caption{The absolute value of the current matrix elements in the eigenstate basis of the Hamiltonian, $|\langle m|\hat J_\mu|n\rangle|$, focusing on the 30 states with lowest energy, for $\mu = x$ in (a) and $\mu = y$ in (b). The dashed lines highlight the target state $|2\rangle$ used in the main text.}
    \label{fig:SM_Fig_1}
\end{figure}

The perturbative analysis above emphasizes the role of the current and kinetic energy operators in the fate of dynamically accessing target states. In the main text, we have focused on the $\hat J_x$ matrix elements in the eigenbasis of the unperturbed Hamiltonian. Now, we extend this analysis by showing the corresponding $y$-direction in Fig.~\ref{fig:SM_Fig_1}(b). Unlike the $x$-direction case, the matrix $\langle m|\hat J_y|n\rangle$ is much sparser and does not exhibit significant matrix elements of the form $\langle 2|\hat J_y|0\rangle$, thus precluding first-order excitation of the target state $|2\rangle$, nor simultaneously large $\langle 2|\hat J_y|m^\prime\rangle$ and $\langle m^\prime|\hat J_y|0\rangle$, preventing relevant second-order current-current excitation terms. This justifies the choice of $x$-polarized pulses we implement in the main text.

\begin{figure}[t!]
    \centering
    \includegraphics[width=1\linewidth]{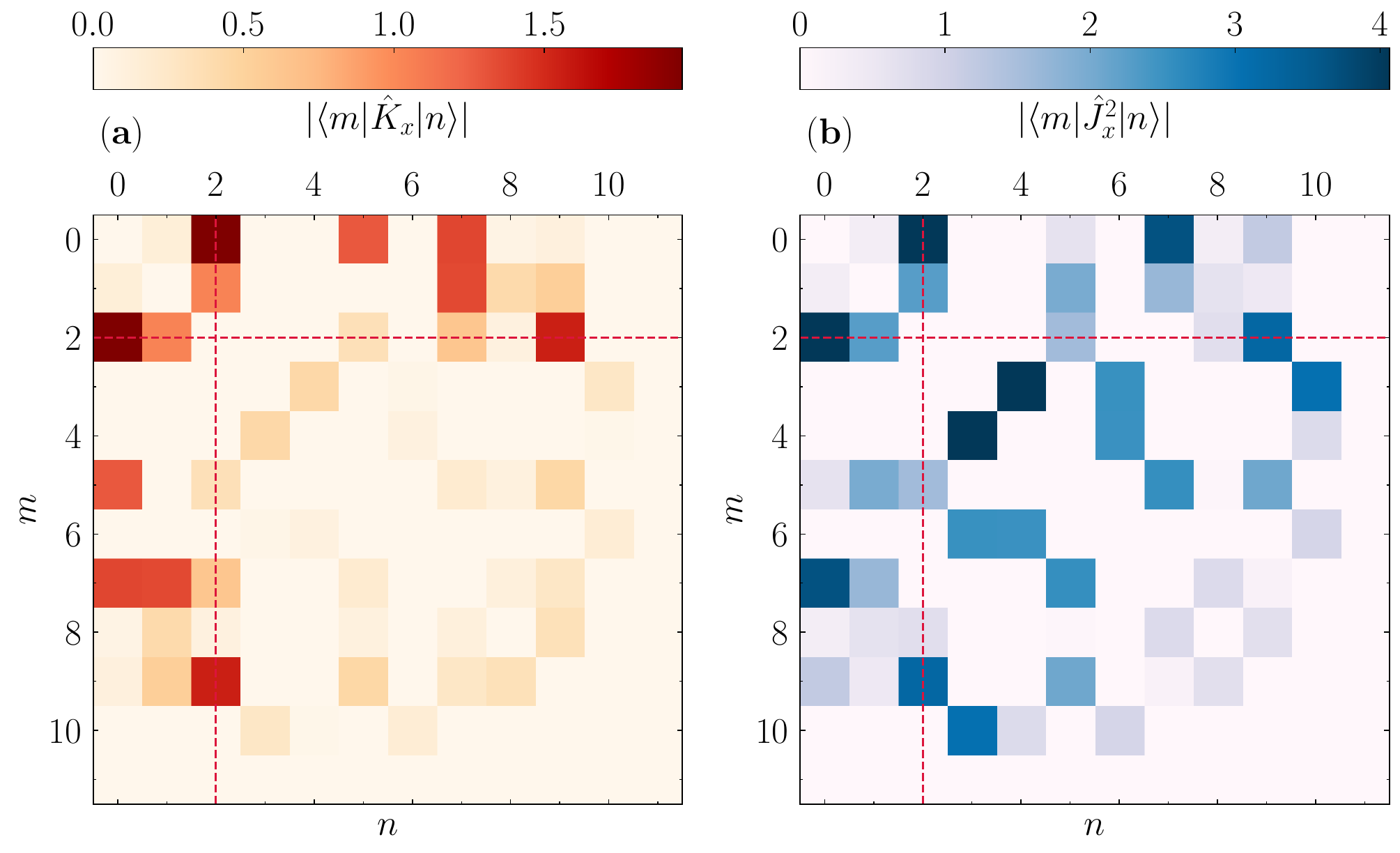}
    \caption{The absolute value of the matrix elements of the $x$-direction kinetic energy (a) and $x$-direction squared-current (b) operators in the eigenstate basis of the Hamiltonian, focusing on the 12 states with lowest energy. Here, the diagonal elements are set to zero for visualization (they are typically much larger). The dashed lines highlight the target state $|2\rangle$ used in the main text.}
    \label{fig:SM_Fig_2}
\end{figure}

Additionally, a strong indication that a second-order process is responsible for the excitation of eigenstate $|2\rangle$ is given by the analysis of the matrix elements of $\hat J_x^2$ in the eigenbasis of $\hat{\cal H}$. As Eqs.~\eqref{eq:second_order_amplitude} and \eqref{eq:second_order_via_pulse} indicate, relevant terms as $\propto\sum_{m^\prime}\langle m|\hat J_x|m^\prime\rangle\langle m^\prime|\hat J_x|0\rangle=\langle m|\hat J_x^2|0\rangle$ need to be large for this type of contribution to be important. Indeed, as Fig.~\ref{fig:SM_Fig_2}(b) shows, the most significant matrix elements of $\hat J_x^2$ in the low-lying spectrum are in between states $|2\rangle$ and $|0\rangle$.

Finally, contributions to a first-order transition between $|0\rangle$ and $|2\rangle$ under the time-dependent perturbation can be obtained using the kinetic energy operator, see Eq.~\eqref{eq:first_order_via_pulse}, with matrix elements proportional to $\langle m|\hat K_x|0\rangle$. As Fig.~\ref{fig:SM_Fig_2}(a) indicates, $|\langle 2|\hat K_x|0\rangle|$ is the largest in the low-lying spectrum. Nonetheless, because the coefficient in the first-order transition amplitude is $A^2(t)$ for the kinetic energy operator, unlike $A(t)$ for the current one, this dictates that such kinetic-driven excitation is not significantly favored for small pulse amplitudes.

\section{Density dynamics}
\label{app:dens_dyn}
\begin{figure}[h!]
    \centering
    \includegraphics[width=1\linewidth]{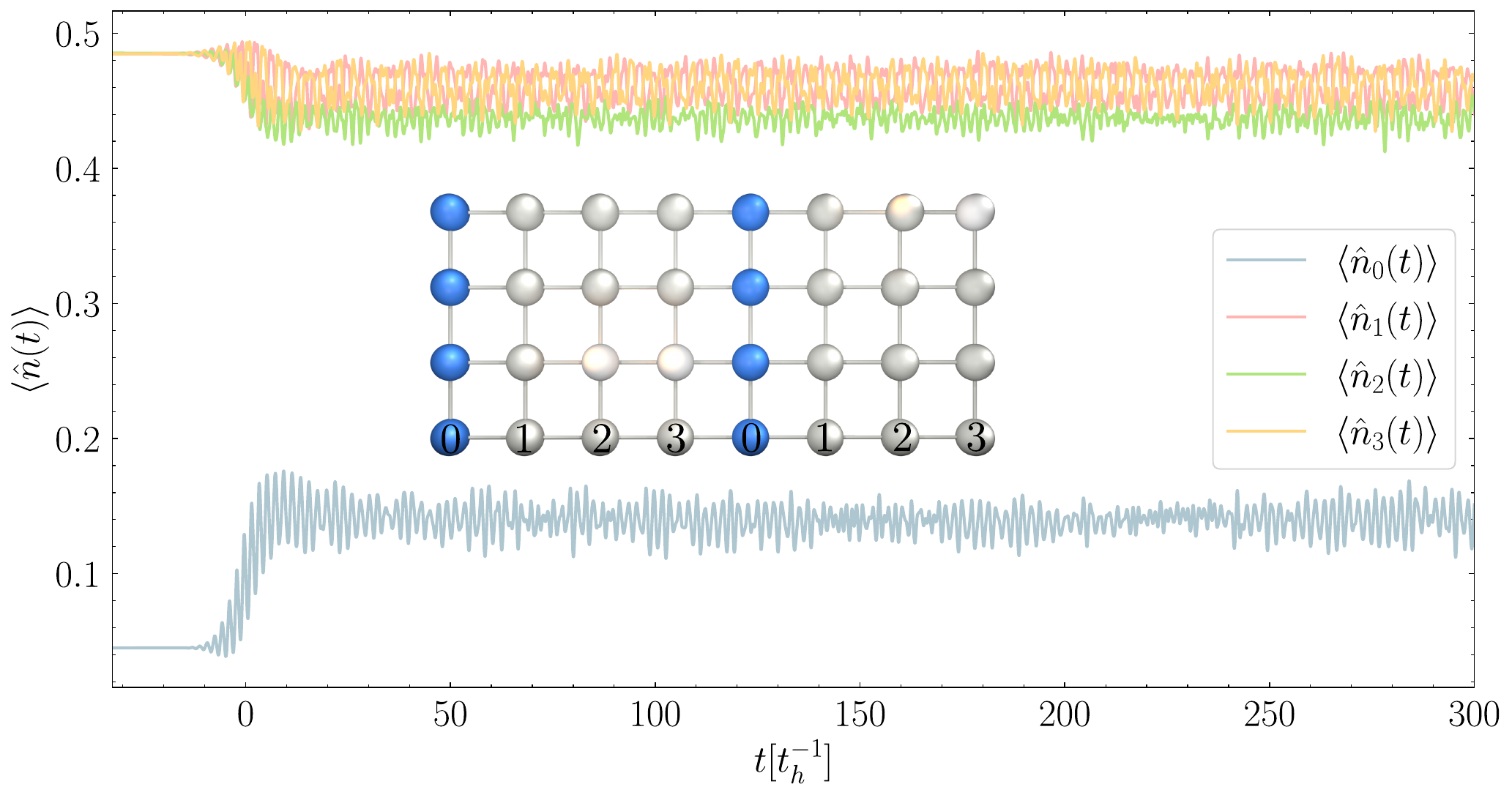}
    \caption{Dynamics for the density resolved in the inequivalent sites under periodic boundary conditions -- inset depicts the lattice with inequivalent sites annotated from 0 to 3. Photoirradiation conditions are the same of Fig.~\ref{fig:fig1} in the main text: $\Omega = 1.713\ t_h$, $A_0 = 0.62$ and $t_d = 6.5 t_h^{-1}$, for Hamiltonian parameters $V/t_h=4$ and $V_0/t_h=5$.}
    \label{fig:SM_Fig_3}
\end{figure}

The main text focuses on the melting of the charge-order reversal under the pulses, releasing the suppressed zero-momentum occupancy. Meanwhile, we can also investigate the fate of the imposed stripes under photoirradiation in terms of the local chemical potential $V_0$. Figure \ref{fig:SM_Fig_3} shows the dynamics of the densities for the four inequivalent sites (under periodic boundary conditions) through photoirradiation with the same conditions as Figs.~\ref{fig:fig1} and \ref{fig:fig3} of the main text, i.e., under maximization of the late-time overlap $|\langle 2|\psi(t_{\rm max})\rangle|$.

Before the pump, the interstripe density is roughly the same (in particular $\langle\hat n_1\rangle=\langle\hat n_3\rangle$ due to symmetry considerations), and much larger than the one for stripe sites ($\langle\hat n_0\rangle \simeq 0.045$). After photoirradiation, although $\pi$-phase shift is now absent, the density imbalance between stripe and interstripe sites is substantially reduced, but still significant, indicating that the externally imposed stripes still survive the dynamical perturbation, albeit with a smaller strength.
\begin{figure}[b!]
    \centering
    \includegraphics[width=0.7\linewidth]{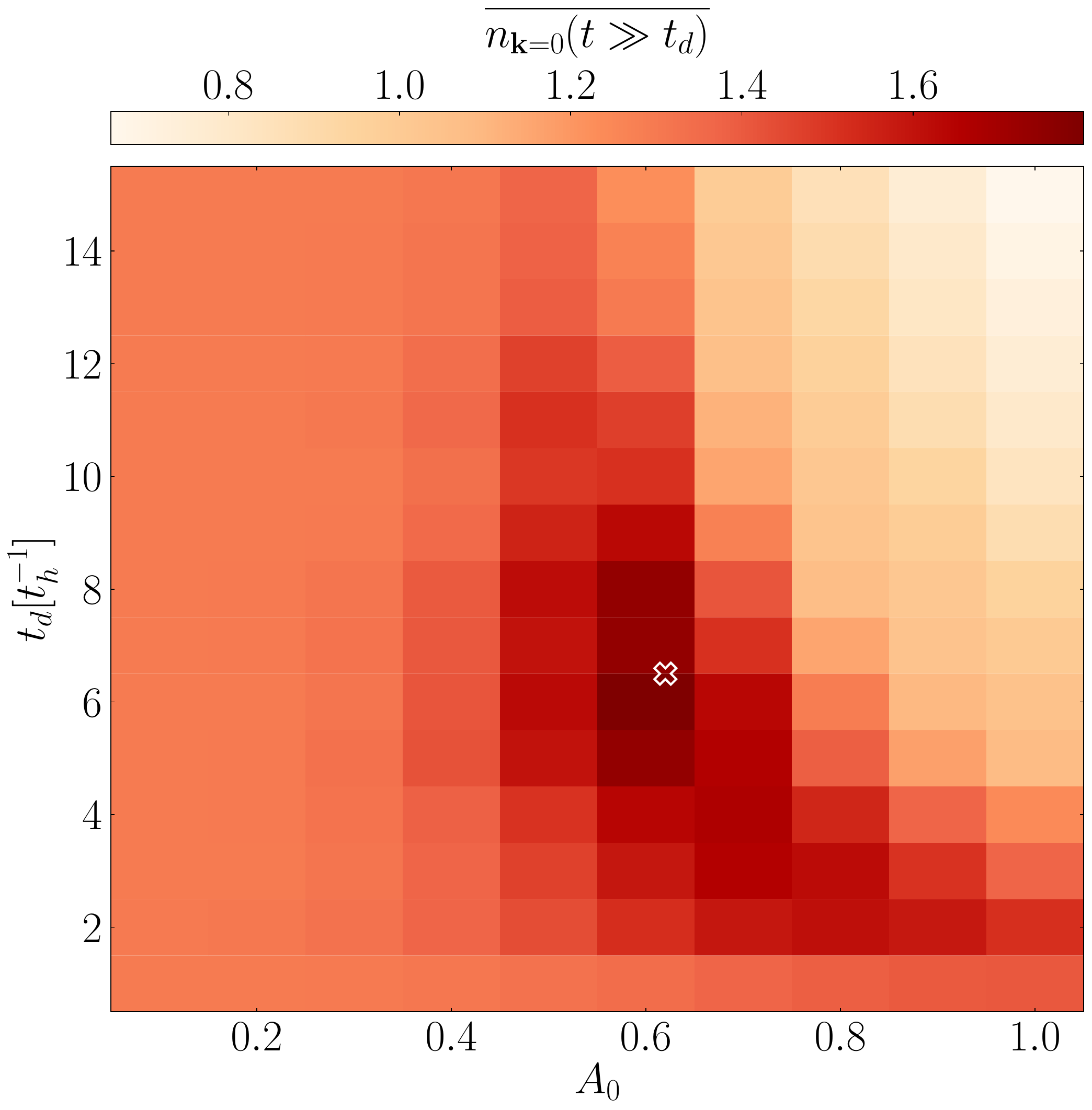}
    \caption{Dependence of the long-time average of the zero momentum occupancy in the pulse parameters $A_0$, pulse amplitude, and $t_d$, pulse width. Here, the pulse frequency is the same as in Figs.~\ref{fig:fig1} and \ref{fig:fig3} of the main text, $\Omega = 1.713\ t_h$. The white-edge marker describes the pair of parameter values ($A_0, t_d$) used in the main text.}
    \label{fig:SM_Fig_4}
\end{figure}
\section{Maximizing zero-momentum occupancy}
\label{app:max_nk0}
In the main text, we establish what pump parameters, namely $A_0$, $t_d$, and $\Omega$,  maximize the overlap with the target state, the equilibrium eigenstate $|2\rangle$ of $\hat {\cal H}$, as emphasized in Fig.~\ref{fig:fig4}(a). We now show that this set of parameters also maximizes the late-time averaged zero momentum occupancy $\langle \hat n_{{\bf k}=0}\rangle$ sufficiently after the photoirradiation. Figure~\ref{fig:SM_Fig_4} shows the time average of the zero momentum occupancy at long times ($t \in[4.5t_d-4t_h^{-1}, 4.5t_d]$) as a function of the pump amplitude and width for the frequency $\Omega =1.713 \ t_h$ --- it is clear that condensation is enhanced in the same range of parameters in which the overlap $|\langle\psi(t\gg t_d)|2\rangle|^2$ is maximized.

\begin{figure}[t!]
    \centering
    \includegraphics[width=1\linewidth]{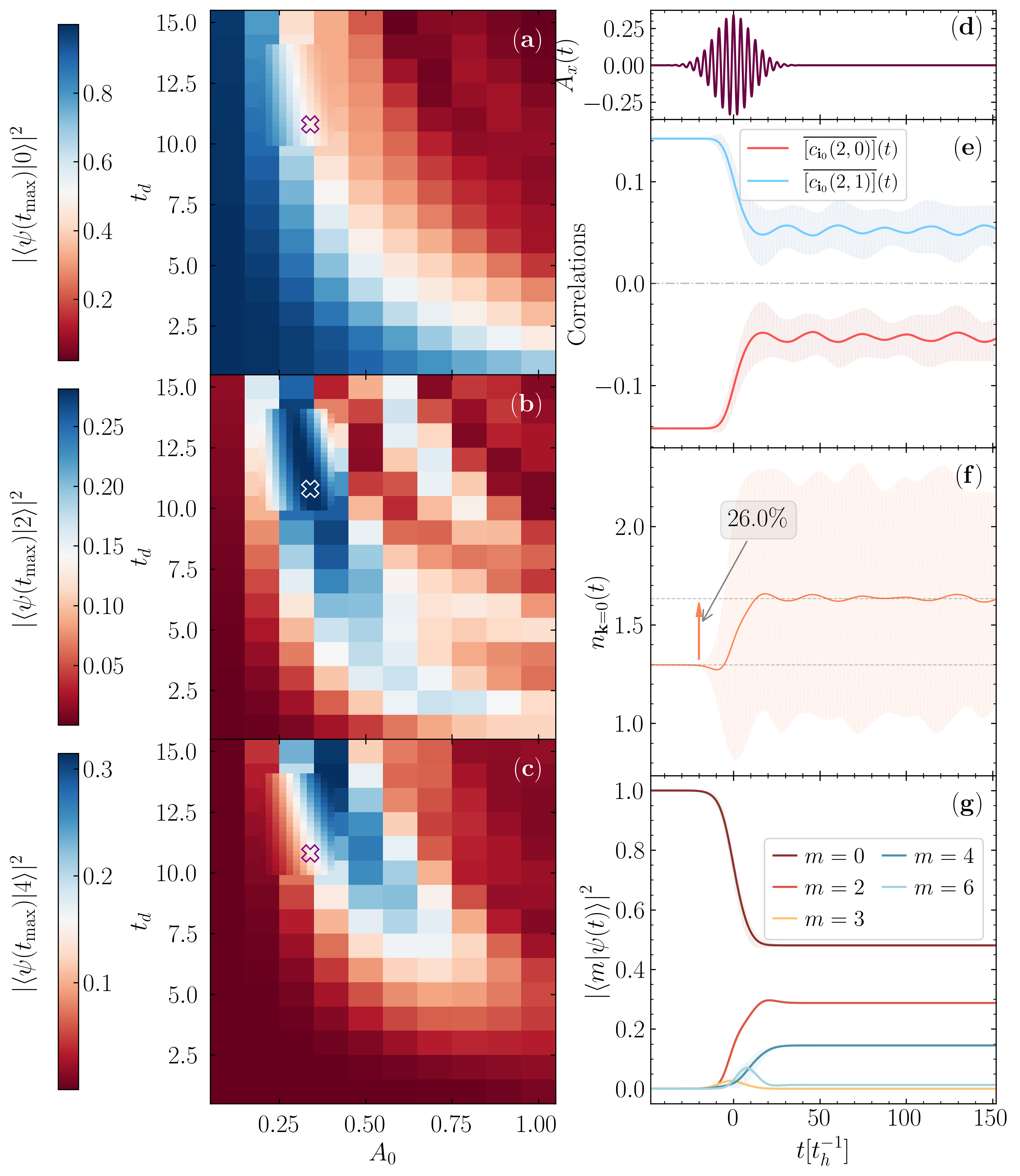}
    \caption{Dependence on $A_0$ and $t_d$ of the long-time weights of equilibrium states in $|\psi(t)\rangle$, for $|m=0\rangle$ (a), $|2\rangle$ (b), and $|4\rangle$ (c) --- the marker assigns the parameters $(A_0, t_d) = (0.34, 10.8)$ resulting in the maximum overlap with the target state $|2\rangle$. (d) Pulse representation for these parameter settings. Dynamics of the connected density correlations across a stripe (e), and the zero momentum occupancy (f). (g) Selected overlaps, by higher values in the dynamics, of the time-evolving state during the photoirradiation process. Data in the plots employ the second-order resonant photoirradiation $\Omega = 1.491\ t_h = \omega_{42}$.}
    \label{fig:SM_Fig_5}
\end{figure}

\section{Other intermediate states}
\label{app:other_interm}
While analyzing the matrix elements of the current operator $\hat J_x$ [Fig.~\ref{fig:SM_Fig_1}], we notice that other states can, in principle, serve as intermediate states for a dynamical transition $|0\rangle\to|2\rangle$. For instance, considering $|m^\prime = 4\rangle$, we notice that $|\langle 2|\hat J_x|4\rangle| = 0.24$ and $|\langle 4|\hat J_x|0\rangle| = 1.73$ --- the largest among the lowest 30 eigenpairs of $\hat {\cal H}$. Thus, using a similar prescription of the main text for the case $|m^\prime = 6\rangle$, resonant and energy conservation conditions lead, respectively, to $ E_4-E_0 \approx n_1\Omega$ and $E_4-E_0 = (n_1+n_2)\Omega$. Subtracting these, we obtain $\omega_{42} \simeq -n_2\Omega$. As in the main text, if we consider the $|4\rangle\to|2\rangle$ dynamical process as a one-photon emission ($n_2=-1$), we can set the pump frequency $\Omega = \omega_{42} \simeq 1.491$. Now combining with the resonant condition, yields $\omega_{40}/\Omega \approx n_1= 4.475/1.491 \simeq 3$, indicating a similar three quantum absorption scenario.

Besides that, additional tuning of the pulse amplitudes and time-width is needed in order to target the desired state, $|2\rangle$, properly. This is accomplished by analyzing the long-time weight of selected equilibrium eigenstates $|m\rangle$ in $|\psi(t)\rangle$, as shown in Fig.~\ref{fig:SM_Fig_5}(a)--(c). Maximimization of $|\langle 2| \psi(t_{\rm max})\rangle|^2$ is achieved, with $\Omega = \omega_{42}=1.491 \ t_h$ at $A_0=0.34$ and $t_d = 10.8\ t_h^{-1}$. Focusing on these parameters [see Fig.~\ref{fig:SM_Fig_5}(d) for the pulse representation], we show the dynamics of both the connected density correlations [Fig.~\ref{fig:SM_Fig_5}(e)] as well as the zero momentum occupancy [Fig.~\ref{fig:SM_Fig_5}(f)]. In this case, the enhancement of the average $n_{{\bf k}=0}$ of about $26\%$ after the pulse is masked by the substantial oscillations that well encompass values below that of the equilibrium regime. Moreover, despite the reduction of the $\pi$-phase density correlations across a stripe, their melting is not achieved in these settings. This suggests that the dynamical addressing of the target state was not optimized.

Indeed, as the dynamics of the overlaps in Fig.~\ref{fig:SM_Fig_5}(g) displays, the weight of $|2\rangle$ in $|\psi(t)\rangle$ is yet small in comparison with the one with the initial state, the ground-state of $|0\rangle$. As a consequence, the targeting is only partial, in contrast to that obtained in the main text.

A critical discussion we left aside in our analysis concerns the system size dependence of our results. As we establish, resonant conditions depend on the low-lying spectrum properties of the relevant Hamiltonians. The issue is that, with larger system sizes, while the energy levels (such as those for the ground state and the target state) are extensive quantities that significantly change with the system size considered, their difference (i.e., the gaps) required to find resonance conditions via the pump has only subextensive corrections. This means that the results could, in principle, be generalized to larger system sizes without the need for dramatic fine-tuning of the pump frequencies when approaching the thermodynamic limit, thereby attesting to the generality of our findings.

\section{Decay of Single-Particle Correlations}
\label{app:LR_corr}
\begin{figure}[t!]
    \centering
    \includegraphics[width=1\linewidth]{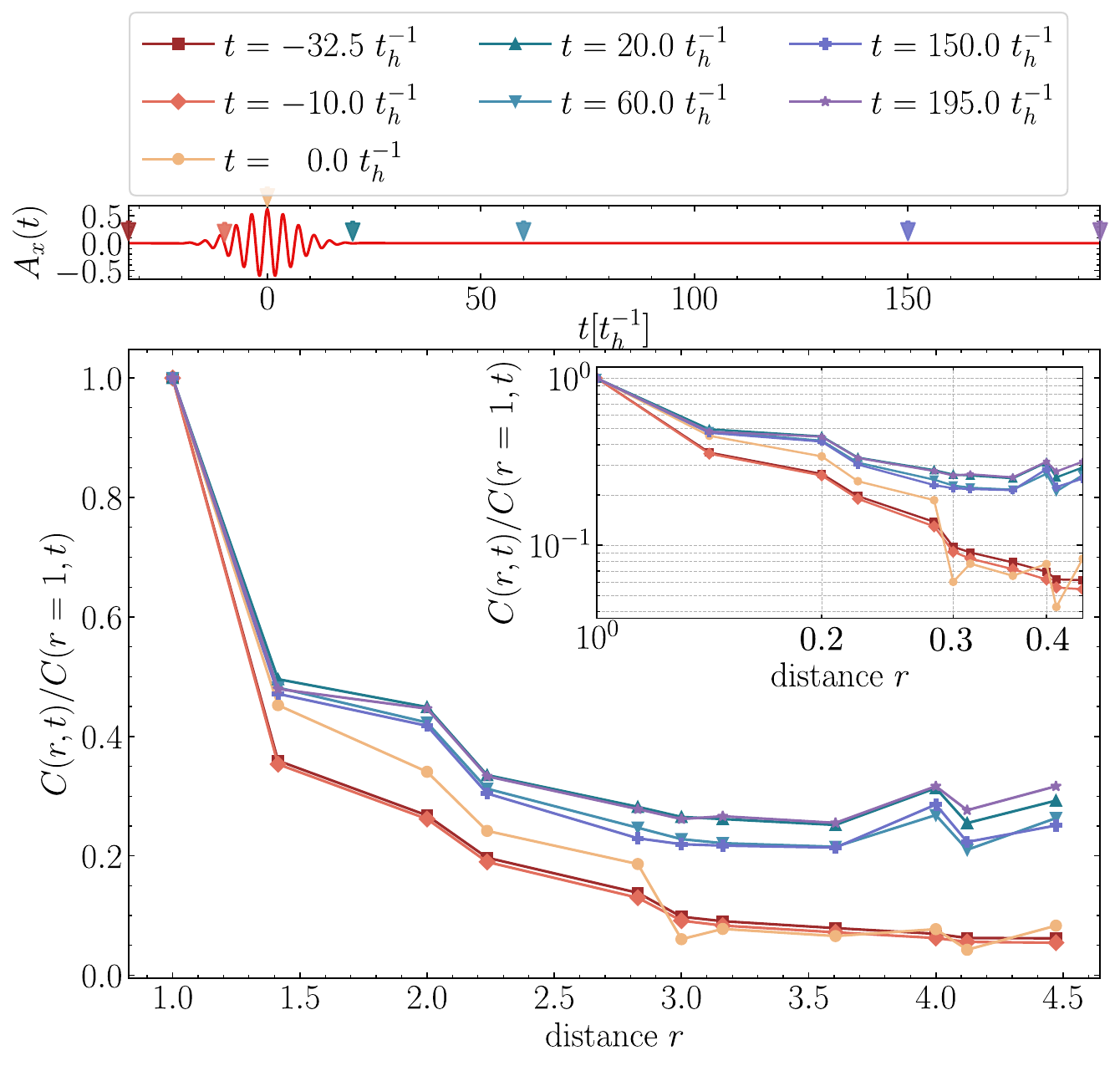}
    \caption{Spatial decay of single-particle correlations $C(r,t)/C(r=1,t)$ at representative times indicated by the vertical arrows in the pump profile (upper panel). Warm (cool) colors correspond to pre- (post-) irradiation states. After the pulse, the correlations saturate at large distances, signaling the emergence of long-range coherence consistent with the enhanced condensate fraction $f_0(t)$ reported in the main text. The inset displays the same data on a log-log scale. Here, the pump parameters are the same as in Fig.~\ref{fig:fig1} of the main text.}
    \label{fig:SM_Fig_6}
\end{figure}

In the main text, we use quantities to probe condensation and phase coherence, the zero-momentum occupancy $\hat n_{{\bf k}=0}$ and the condensate fraction $f_0$, which inherently build on top of the single-particle correlations, forming the single-particle density matrix (SPDM) correlations,
\begin{equation}
C(r,t) = \langle \hat b_{{\bf i}}^{\dagger}(t)\hat b_{\bf i+r}^{\phantom{\dagger}}(t)\rangle\ .
\end{equation}
To understand how the enhanced condensation is manifested microscopically in the lattice, Fig.~\ref{fig:SM_Fig_6} presents the spatial decay of the SPDM correlations averaged over all site pairs separated by a distance $r=||{\bf r}||$, in a translationally invariant fashion. The correlations are further normalized by their nearest-neighbor value, $C(r,t)/C(r=1,t)$. The upper panel of Fig.~\ref{fig:SM_Fig_6} shows the pump profile $A_x(t)$, with vertical arrows marking the time stamps for which the correlation functions are displayed below. Warm colors (reds and oranges) correspond to times prior to the pulse, while cool colors (blues and purples) denote post-irradiation time-stamps. The lower panel plots $C(r,t)/C(r=1,t)$ as a function of distance, and the inset displays the same data on log-log axes to highlight the emergence of algebraic-like decay.

The crucial and interesting point is that, at long distances, the correlations at late time ($t>0$) do not follow a simple power-law behavior. Instead, $C(r,t)/C(r=1,t)$ tends to saturate toward a finite value, signaling the emergence of a nonvanishing off-diagonal component of the single-particle density matrix. This plateau indicates that the coherence length $\xi(t)$ exceeds the system size, corresponding to the development of global phase coherence and long-range order. In contrast, before the pulse ($t<0$), the correlations decay rapidly, consistent with a short-range correlated stripe state. The observed saturation at long times thus represents the finite-size manifestation of photoinduced condensation, in agreement with the enhanced condensate fraction $f_0(t)$ and Drude weight reported in the main text.

\section{Distinction between Drude and superfluid weights}
\label{app:D_s_out_of_eq}
In the main text, we introduce a protocol for the out-of-equilibrium determination of the Drude weight $D$, based on the insertion of a probe linear-in-time vector potential much after the photoirradiation, when the system is closely relaxing toward equilibrium conditions. We now further justify this approach and contrast it with one that enables the out-of-equilibrium determination of a metric of phase coherence, namely the superfluid weight $D_s$.

We start by reviewing the conditions under which one extracts the equilibrium $D$ and $D_s$. In equilibrium, the current response to a $x$-oriented vector potential $A_x$ can be expressed as~\cite{Scalapino1993}
\begin{equation}
    \langle \hat J_x\rangle = -\langle \hat K_x\rangle A_x + \langle \hat \Pi_{xx}({\bf q}, \omega)\rangle A_x\, ,
\end{equation}
where $\langle \hat K_x\rangle$ and $\langle \hat\Pi_{xx}\rangle$ are the diamagnetic and paramagnetic contributions, respectively, the latter mapping the current-current correlator. Two distinct stiffnesses arise depending on the order of limits taken for $\omega$ and ${\bf q}$:
\begin{align}
D^{(x)} &= \langle -\hat K_x\rangle - \lim_{\omega \to 0} \langle \hat \Pi_{xx}({\bf q}{=}0,\omega)\rangle , \notag \\
D_s^{(x)} & = \langle -\hat K_x\rangle - \lim_{q_x=0,q_y \to 0} \langle \hat \Pi_{xx}({\bf q},\omega{=}0)\rangle .
\end{align}
The first quantity, $D^{(x)}$, is the equilibrium Drude weight or charge stiffness, governing the zero-frequency weight of the optical conductivity and quantifying ballistic charge transport in the long-wavelength (${\bf q}=0$) limit. The second, $D_s^{(x)}$, is the equilibrium superfluid weight, obtained from the static ($\omega=0$) transverse limit ($q_x=0,q_y\!\to\!0$) and associated with phase rigidity and the Meissner response.

An equivalent, and often more transparent, way to express these stiffnesses is in terms of the curvature of the total energy with respect to a uniform twist $\Phi_x = A_x L_x$ applied along the periodic direction~\cite{Kohn1964,Scalapino1993}. For the ground state $|m=0\rangle$ with energy $E_0(\Phi_x)$, one can write
\begin{equation}
    \frac{D^{(x)}}{\pi} = \frac{L_x^2}{L_xL_y}\left.\frac{\partial^2 E_0(\Phi_x)}{\partial \Phi_x^2}\right|_{\Phi_x=0}\, .
\end{equation}
In finite systems, $E_0(\Phi_x)$ generally exhibits cusps whenever the ground-state wave function crosses another state with a different current quantum number. As discussed in Ref.~\cite{Scalapino1993}, in one dimension, such crossings remain well-separated even in the thermodynamic limit, allowing one to adiabatically follow the ground state and unambiguously evaluate the curvature of $E_0(\Phi_x)$ at $\Phi_x=0$, which yields the Drude weight. In higher dimensions, however, the crossings proliferate and become dense, so that the ground-state energy as a function of twist is instead represented by the smooth envelope of the many-body branches $E_n(\Phi_x)$. In this case, computing $\partial^2 E/\partial\Phi_x^2$ at fixed system size and extrapolating gives the adiabatic curvature associated with the Drude weight $D$, whereas taking the thermodynamic limit first and differentiating the envelope gives the superfluid weight $D_s$. The two quantities coincide in one-dimensional systems, but differ in higher dimensions or whenever level crossings destroy adiabatic continuity.

These definitions, expressed in terms of energy derivatives, are fully equivalent to those from the current response, provided the same limit ordering is respected. In equilibrium, the distinction between the Drude and superfluid weights stems from the different order in which temporal and spatial limits are taken: taking $\omega \to 0$ before ${\bf q} \to 0$ (the long wavelength limit) yields the Drude weight $D$, while reversing the order (the static transverse case) gives the superfluid weight $D_s$. This distinction can be dynamically mirrored by preparing nonequilibrium states that realize, respectively, a uniform time-dependent or a static spatially twisted gauge field.

To make this correspondence explicit, recall that in a gauge where the scalar potential is zero, the electric field follows from the time derivative of the vector potential, $E_x(t) = -\partial_t A_x(t)$. A slowly varying, linear-in-time ramp of $A_x(t)$ therefore produces a uniform, nearly static electric field and probes the small-frequency, spatially uniform response, i.e., the ${\bf q}=0$, $\omega\!\to\!0$ limit relevant to charge transport. In contrast, a static vector potential $\delta A_x$ has $\partial_t A_x = 0$ and thus no electric field. Its effect enters solely through the spatial boundary conditions, or equivalently, through a uniform twist $\Phi_x = A_x L_x$ in the many-body wavefunction, which probes the system's response to a static phase gradient. This corresponds to the static ($\omega=0$), long-wavelength (${\bf q}\!\to\!0$) limit that underlies the equilibrium definition of the superfluid weight.
\begin{figure}[t!]
    \centering
    \includegraphics[width=1\linewidth]{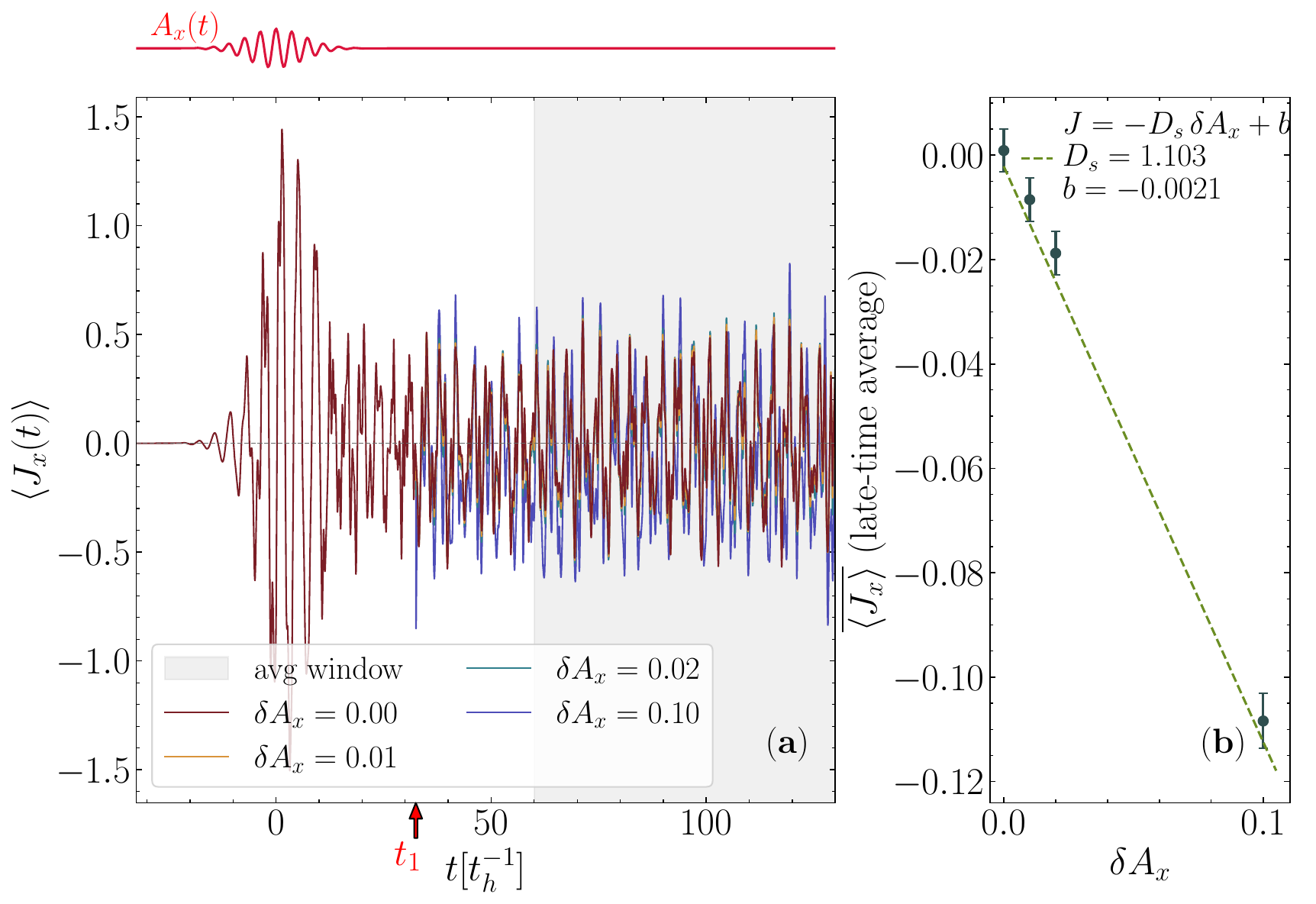}
    \caption{(a) Time evolution of the current $\langle J_x(t)\rangle$ after static vector-potential quenches $A_x=\delta A_x$ applied at $t_1=5t_d$. The original photoirradiation pulse, centered at $t_0=0$, is shown at the top. (b) Late-time average $\overline{\langle J_x\rangle}$ [time-window given by the shaded region in (a)] versus $\delta A_x$, whose linear slope determines $D_s^{(\mathrm{eff}, x)}=-\partial J_x/\partial A_x|_{A_x\to0}$. Dashed line displays a linear fit to the data, fitting parameters in the legend, whereas error bars are the standard error of the mean of $\langle J_x(t)\rangle$ within the averaging window. Pump parameters are the same as Fig.~\ref{fig:fig1} of the main text, and of Fig.~\ref{fig:SM_Fig_6}.}
    \label{fig:SM_Fig_7}
\end{figure}
In this way, the ramp and static protocols directly implement, in real time, the two complementary limits that distinguish $D$ and $D_s$ in equilibrium. The former probes the inertial response to an applied electric field (longitudinal conductivity), while the latter probes the phase rigidity of the current in a time-independent twist (transverse superfluid response). The resulting persistent current encodes the slope of the smooth envelope of the energy with respect to the twist, \cite{Scalapino1993}
\begin{equation}
    D_s^{({\rm eff},x)} = -\frac{\partial J_x}{\partial A_x}\Big|_{A_x\to0}\, ,
\end{equation}
providing a nonequilibrium analogue of the superfluid weight.

Thus, to quantify the out-of-equilibrium superfluid response, we promote vector-potential quenches $A_x = \delta A_x$ at times $t_1 = 5t_d$, where the pulse has significantly faded. Computing the current response $\langle J_x(t)\rangle$ for various $\delta A_x$ [Fig.~\ref{fig:SM_Fig_7}(a)] allows one to extract $D_s^{(\mathrm{eff}, x)}$ from the slope of the steady-state current $\overline{\langle J_x\rangle}$ versus $\delta A_x$ [Fig.~\ref{fig:SM_Fig_7}(b)]. The relation $J_x = -D_s^{(x)} A_x$ implies that a positive superfluid stiffness corresponds to a restoring current flowing opposite to the imposed vector potential. The simultaneous finite values of $D^{(x)}$ (see main text) and $D_s^{(x)}$ thus demonstrate that photoirradiation induces not only ballistic charge transport but also a phase-coherent superfluid response.

\section{Symmetries and targeting of states}
\label{app:symmetries}
In the above discussion, we explore how time-dependent perturbation theory provides a framework for understanding which states can be targeted via matrix elements of relevant operators, including the current $\hat J_x$. In particular, because this operator is odd under reflection symmetry, a selection rule emerges, wherein only eigenstates with opposite parities may possess a finite matrix element, i.e., $\langle {\rm odd}|\hat J_x|{\rm even}\rangle\neq 0$, whereas $\langle {\rm even}|\hat J_x|{\rm even}\rangle$ and $ \langle {\rm odd}|\hat J_x|{\rm odd}\rangle$ are identically zero.

This can be seen by considering the reflection operator $\hat {\cal R}_3$, a reflection about the stripe located at $x=3$ [see Fig.~\ref{fig:fig1}(a)], which, combined with the periodic boundary conditions, leads to the mapping of the site coordinates $\hat {\cal R}_3: (x,y) \mapsto (-x \mod L_x, y)$. If labeling the sites by their $x$-coordinate in our $L_x = 8$ ladder, this means this map preserves the stripe sites, $3 \mapsto 3$ and $7 \mapsto 7$, while symmetric sites about a stripe are mapped to one another: $2 \mapsto 4$, $1 \mapsto 5$, and $0\mapsto 6$. Since $[\hat {\cal R}_3, \hat {\cal H}]=0$, the eigenstates $|m\rangle$ of $\hat {\cal H}$ are also eigenstates of $\hat {\cal R}_3$ and possess a definite parity, $\hat {\cal R}_3|m\rangle=\pm|m\rangle$.

Turning back to the symmetry of the current operator, one sees that it is antisymmetric under the reflection $\hat{\cal R}_3$, i.e.,
$\hat{\cal R}_3 \hat J_x \hat{\cal R}_3^{-1} = -\hat J_x$.
As a consequence, the matrix elements obey
\[
\langle m|\hat J_x|n\rangle = -(\pm_m)(\pm_n)\langle m|\hat J_x|n\rangle ,
\]
which can only be satisfied if the states $|m\rangle$ and $|n\rangle$ possess opposite parities. Inspecting Fig.~\ref{fig:fig4}(c), one therefore finds that the \emph{even}-parity ground state $|0\rangle$ has finite current matrix elements only with \emph{odd}-parity excited states, namely $|3\rangle$, $|4\rangle$, $|6\rangle$, and $|10\rangle$. By contrast, the remaining low-lying states $|1\rangle$, $|2\rangle$, $|5\rangle$, $|7\rangle$, $|8\rangle$, and $|9\rangle$ are even-parity, as evidenced by their finite overlap with $|0\rangle$ under the reflection-symmetric operator $\hat J_x^2$, see Fig.~\ref{fig:SM_Fig_2}(b).

These symmetry considerations directly inform the structure of the photoexcitation pathways. Because $\hat J_x$ is odd under $\hat{\cal R}_3$, odd-parity excited states are directly accessible from the even-parity ground state via single-photon processes. In contrast, even-parity excited states cannot be reached at leading order and instead require higher-order processes involving even powers of $\hat J_x$, such as $\hat J_x^2$, which are reflection-symmetric. Consequently, at small pump amplitudes the dynamics is dominated by odd-parity states, while increasing the pump strength enables multiphoton processes that populate even-parity states through two-photon resonances and higher odd-parity states through three-photon pathways.

This structure explains the observed hierarchy of state populations under the drive. At small pump amplitudes, the dynamics is dominated by odd-parity states such as $|3\rangle$, which couple linearly to the even-parity ground state via single-photon processes mediated by $\hat J_x$. As the pump amplitude increases, higher-order processes become effective, allowing access to even-parity states such as $|2\rangle$ through two-photon resonances, as well as to higher odd-parity states such as $|6\rangle$ via three-photon pathways. The enhanced occupation of these states under a drive frequency $\Omega\simeq1.713,t_h$ is therefore consistent with both their symmetry under reflection and their energetic proximity to the corresponding multiphoton resonance conditions.

We emphasize, however, that reflection symmetry alone does not determine the efficiency of state targeting. While symmetry dictates which matrix elements are allowed, the actual population transfer depends quantitatively on energy denominators, matrix-element magnitudes, and the pump amplitude. In particular, once the pump is applied, the time-dependent Hamiltonian explicitly breaks $\hat{\cal R}_3$, and parity ceases to be a conserved quantity during the dynamics. The observed state selectivity, therefore, reflects a combination of symmetry-based selection rules and dynamical resonance effects, both of which are explicitly captured in our numerical time-evolution and perturbative analyses.

\bibliography{Reference}

\end{document}